\documentclass[traditabstract]{aa} % for the abstract without structuration 
\usepackage{graphicx}
\usepackage{marginnote}
\usepackage{txfonts}
\usepackage{subfigure}
\usepackage{natbib}
\usepackage{amstext}
\usepackage[verbose]{placeins}
\usepackage{tikz}
\usepackage{array}
\usepackage{amssymb}
\usepackage{xcolor}
\usepackage{multirow}
\usepackage{longtable}
\usepackage{pdflscape}
\usepackage{color, colortbl}
\definecolor{Gray}{gray}{0.9}
\usepackage{hyperref}

\newcommand{\bd}{\begin{displaymath}}
\newcommand{\ed}{\end{displaymath}}
\newcommand{\be}{\begin{equation}}
\newcommand{\ee}{\end{equation}}
\newcommand{\beaa}{\begin{eqnarray*}}
\newcommand{\eeaa}{\end{eqnarray*}}
\newcommand{\bea}{\begin{eqnarray}}
\newcommand{\eea}{\end{eqnarray}}

\def\GLEE{\textsc{Glee}}
\def\GLaD{\textsc{Glad}}
\def\GG{\GLEE \, \& \GLaD}
\def\Pyautolens{\textsc{PyAutoLens}}

\defcitealias{schuldt21a}{S21b}

\begin{document}

   \title{HOLISMOKES - IX. Neural network inference of strong-lens parameters and uncertainties from ground-based images\thanks{The network code is available under \url{https://github.com/shsuyu/HOLISMOKES-public/tree/main/HOLISMOKES_IX}}}

   \titlerunning{HOLISMOKES - IX. Neural network inference of strong-lens parameters and uncertainties}

   \author{S.~Schuldt\inst{1}\inst{,2}
     \and
     R.~Ca\~{n}ameras\inst{1}
     \and
     Y.~Shu\inst{1}\inst{,3}
     \and
     S.~H.~Suyu\inst{1}\inst{,2}\inst{,4}
     \and
     S.~Taubenberger\inst{1}
     \and
     T.~Meinhardt\inst{5}
     \and
     L.~Leal-Taix\'{e}\inst{5}
          }

   \institute{Max-Planck-Institut f\"ur Astrophysik, Karl-Schwarzschild Str.~1, 85748 Garching, Germany \\
              \email{schuldt@mpa-garching.mpg.de}
              \and  
              Technical University of Munich, TUM School of Natural Sciences, Department of Physics, James-Franck-Str.~1, 85748 Garching, Germany
              \and
              Ruhr University Bochum, Faculty of Physics and Astronomy, Astronomical Institute (AIRUB), German Centre for Cosmological Lensing, 44780 Bochum, Germany
              \and
              Academia Sinica Institute of Astronomy and Astrophysics (ASIAA), 11F of ASMAB, No.1, Section 4, Roosevelt Road, Taipei 10617, Taiwan
              \and
              Technische Universit\"at M\"unchen, Dynamic Vision and Learning Group, Bolzmannstr. 3, 85748 Garching, Germany
             }

   \date{Received --; accepted --}

% \abstract{}{}{}{}{} 
% 5 {} token are mandatory

  \abstract
  % context heading (optional)
  % {} leave it empty if necessary  
      {Modeling of strong gravitational lenses is a necessity for further applications in astrophysics and cosmology. With the large number of detections in current and upcoming surveys, such as the Rubin Legacy Survey of Space and Time (LSST), it is pertinent to investigate automated and fast analysis techniques beyond the traditional and time-consuming Markov chain Monte Carlo sampling methods. Building upon our (simple) convolutional neural network (CNN), we present here another CNN, specifically a residual neural network (ResNet), that predicts the five mass parameters of a singular isothermal ellipsoid (SIE) profile (lens center $x$ and $y$, ellipticity $e_\text{x}$ and $e_\text{y}$, Einstein radius $\theta_\text{E}$) and the external shear ($\gamma_\text{ext,1}$, $\gamma_\text{ext,2}$) from ground-based imaging data. In contrast to our previous CNN, this ResNet further predicts the 1$\sigma$ uncertainty for each parameter. To train our network, we use our improved pipeline to simulate lens images using real images of galaxies from the Hyper Suprime-Cam Survey (HSC) and from the Hubble Ultra Deep Field as lens galaxies and background sources, respectively. We find very good recoveries overall for the SIE parameters, especially for the lens center in comparison to our previous CNN, while significant differences remain in predicting the external shear. From our multiple tests, it appears that most likely the low ground-based image resolution is the limiting factor in predicting the external shear. Given the run time of milli-seconds per system, our network is perfectly suited to quickly predict the next appearing image and time delays of lensed transients. Therefore, we use the network-predicted mass model to estimate these quantities and compare to those values obtained from our simulations. Unfortunately, the achieved precision allows only a first-order estimate of time delays on real lens systems and requires further refinement through follow-up modeling. Nonetheless, our ResNet is able to predict the SIE and shear parameter values in fractions of a second on a single CPU, meaning that we are able to efficiently process the huge amount of galaxy-scale lenses  expected in the near future.
        }

  % aims heading (mandatory)
  % {}
  % methods heading (mandatory)
  % {}
  % results heading (mandatory)
  % {}
  % conclusions heading (optional), leave it empty if necessary 
  % {}

\maketitle

%________________________________________________________________
\section{Introduction}
\label{sec:introduction}

% Into similar to HOLISMOKES IV

Gravitational lensing has become a very powerful tool in astrophysics, especially in combination with others, such as the lens velocity dispersion measurements \citep[e.g.,][]{barnabe11, barnabe12, yildirim20} and the galaxy rotation curves \citep[e.g.,][]{strigari13, hashim14}, which both help to probe the mass structure of galaxies. In particular, gravitational lensing allows us to measure the total mass of the lens \citep{dye05, treu10} and thus to study the fraction and distribution of dark matter \citep[DM; e.g.,][]{schuldt19, baes21, shajib21, wang22}, as well as its nature \citep[e.g.,][]{basak22, gilman21}. Moreover, one can use lensing to study high-redshift sources thanks to the lensing magnification \citep[e.g.,][]{dye18, lemon18, mcgreer18, rubin18, salmon18, shu18} by reconstructing the surface brightness of the source, which can be used to reveal information about the evolution of galaxies at higher redshifts \citep[e.g.,][]{warren03, suyu06a, nightingale18, rizzo18, chirivi20}.

In special cases where a variable source, such as a quasar \citep[e.g.,][]{lemon17, lemon18, lemon19, ducourant19, khramtsov19, chan20, chao20, chao21} or supernova (SN), is present as a background object \citep{kelly15, goobar17, rodney21}, one can measure the time delays \citep[e.g.,][]{millon20a, millon20b, huber22} and then constrain the Hubble constant $H_0$ \citep[e.g.,][]{refsdal64, chen19, rusu20, wong20, shajib20, shajib22}, helping to clarify the current discrepancy between the cosmic microwave background (CMB) measurements \citep{planck20} and measurements using the local distance ladder \citep[e.g., the SH0ES project,][]{riess19, riess21}. Furthermore, if one detects a lensed SN on the first image, and predicts the location and time for the next appearing image(s), the SN can be observed at earlier phases than without lensing, which would help to shed light on open questions regarding the SN progenitor system(s).

For this reason, and also because such lensing systems with time-variable background sources are very rare, great effort has been made in recent years to run several large and dedicated surveys, such as the Sloan Lens ACS (SLACS) survey \citep{bolton06, shu17}, the CFHTLS Strong Lensing Legacy Survey \citep[SL2S;][]{cabanac07, sonnenfeld15}, the Sloan WFC Edge-on Late-type Lens Survey \citep[SWELLS;][]{treu11}, the BOSS Emission-Line Lens Survey \citep[BELLS;][]{brownstein12, shu16a, cornachione18}, and the Survey of Gravitationally-lensed Objects in HSC Imaging \citep[SuGOHI;][]{sonnenfeld18a, wong18, chan20, jaelani20a}. In addition to that, several teams used the imaging data from large surveys, such as the Dark Energy Survey \citep[DES; e.g.,][]{jacobs19, rojas22}, the Panoramic Survey Telescope and Rapid Response System \citep[Pan-STARRS; e.g.,][]{lemon18, canameras20}, the Kilo Degree Survey \citep[KiDS; e.g.,][]{petrillo17, petrillo19a, li20, li21c}, and the surveys with the Hyper Suprime-Cam \citep[HSC; e.g.,][]{canameras21b, shu22, jaelani22}, to identify additional lens systems. In total, a few hundred spectroscopically confirmed lenses and many more promising lens candidates have been found so far, but the sample of lens candidates is expected to grow by a factor of around 20 \citep{collett15} with upcoming surveys, such as the Rubin Observatory Legacy Survey of Space and Time \citep[LSST,][]{ivezic08}, which will observe around $20,000$ $ \text{deg}^2$ of the southern hemisphere in six different filters $(u, g, r, i, z, y)$, or the Euclid imaging survey operated by the European Space Agency \citep[ESA;][]{laureijs11}.

To find those strong galaxy--galaxy scale lenses, huge efforts are currently being made to develop fast and automated algorithms in order to classify billions of observed galaxies. There are different methods available, such as geometrical quantification \citep{bom17, seidel07}, spectroscopic analysis \citep{baron17, ostrovski17}, an Arcfinder including color cuts \citep{gavazzi14, maturi14}, or machine learning techniques \citep[e.g.,][]{jacobs17, petrillo17, lanusse18, schaefer18, metcalf19, canameras20, canameras21b, canameras22a, huang20, rojas22, savary22, jaelani22, shu22}.

After detecting the lens candidates, a model describing their total mass distribution is required for nearly all applications; this also helps to reject some false-positive candidates \citep{marshall09, sonnenfeld13, chan15, taubenberger22}. As the sample of known lenses is increasing rapidly, current Monte-Carlo Markov-Chain (MCMC)-based techniques \citep[e.g.,][]{jullo07, suyu10a, sciortino20, fowlie20} are no longer sufficient to model all of them, because the MCMC sampling is very time consuming and resource dependent, and requires a lot of human input. \citet{hezaveh17} therefore proposed and demonstrated the feasibility of using convolutional neural networks (CNNs) to predict the mass model parameter values for high-resolution and pre-processed, lens-light-subtracted images. This was further improved and explored by the same team \citep{levasseur17, morningstar18, morningstar19}, and \citet{pearson19, pearson21} also presented a CNN for modeling high-resolution lens images. For the LSST Dark Energy Science Collaboration, \citet{wagnercarena21} developed so-called Bayesian neural networks to model HST-like lenses after lens light subtraction in analogy to \citet{hezaveh17}, while in their follow-up work, they extended their approach to HST-like images without prior lens light subtraction \citep{park21a}. To complement this, as part of our ongoing Highly Optimized Lensing Investigations of Supernovae, Microlensing Objects, and Kinematics of Ellipticals and Spirals \citep[HOLISMOKES, ][]{suyu20} programme, in \citet[][hereafter S21b]{schuldt21a} we presented a CNN designed to model strongly lensed galaxy images. This CNN predicts the five singular isothermal ellipsoid (SIE) mass model parameter values (lens center $x$ and $y$, complex ellipticity $e_\text{x}$ and $e_\text{y}$, and the Einstein radius $\theta_\text{E}$) for the lens galaxy mass distribution using ground-based HSC images.

In the presented work, which builds upon \citetalias{schuldt21a}, we adopt an SIE profile plus external shear component and include an uncertainty prediction for each parameter, which has already been demonstrated on high-resolution HST-like images \citep{levasseur17, park21a, wagnercarena21}. As this task is much more complex than that in \citetalias{schuldt21a}, it requires a deeper network to cope with the small distortions of the external shear, such that we rely now on residual blocks in the network architecture \citep[ResNet, ][]{he16a}. Moreover, we improved our simulation pipeline, which uses real observed galaxy images as well as corresponding redshift and velocity measurements \citepalias{schuldt21a}, to obtain even more realistic training data. In detail, we add Poisson noise on the simulated arcs, add the external shear component, and improve the automated computation of the first and second brightness moments to obtain the lens light center and ellipticity. Our mock images of lenses also contain other galaxies in the image cutout, allowing our ResNet to cope with realistic lens systems that often have other nearby objects along the line of sight. While we use only simulated data to test the network performance in this work, we apply this network to a smaller sample of 31 real HSC lenses in a companion paper \citep{schuldt22b} and compare the obtained models to the traditional ones obtained through MCMC sampling. The good agreement found there shows that our mock images are realistic and that our demonstrated network performance is trustworthy.

As one of the main advantages of our ResNet is the computational speed, it is perfectly suited to predict the lens mass model parameters of a lensing system with a transient host galaxy as the background source, which can then be used to predict the next appearing image(s) and time delay(s) of the transient. Therefore, in analogy to \citetalias{schuldt21a}, we compare the predicted image positions and time delays using (1) the ground-truth mass model from our simulation pipeline and (2) the predicted model from the network.

The outline of the paper is as follows. We summarize in Sect.~\ref{sec:simulation} the simulation of our training data and changes compared to \citetalias{schuldt21a}. In Sect.~\ref{sec:network}, we describe our network architectures and in Sect.~\ref{sec:results} we present our results. The dedicated tests for the network are summarized in Sect.~\ref{sec:tests}. Our comparisons of the image positions and time delays are reported in Sect.~\ref{sec:ImPosTimeDelays}, before we present our conclusions in Sect.~\ref{sec:conclusion}.

Throughout this work, we assume a flat $\Lambda$CDM cosmology with a Hubble constant $H_0 = 72\, \text{km}\, \text{s}^{-1}\, \text{Mpc}^{-1}$ \citep{bonvin17} and $\Omega_\text{M} =1 -\Omega_\Lambda = 0.32 $ \citep{planck20} in analogy to \citetalias{schuldt21a}.

\FloatBarrier
\section{Simulation of mock lenses}
\label{sec:simulation}

% Short recap of HOLISMOKES IV data/procedure and now differences
% introduce external shear

\begin{figure*}
  \begin{center}
    \includegraphics[trim=0 0 0 0, clip, width=0.7\textwidth]{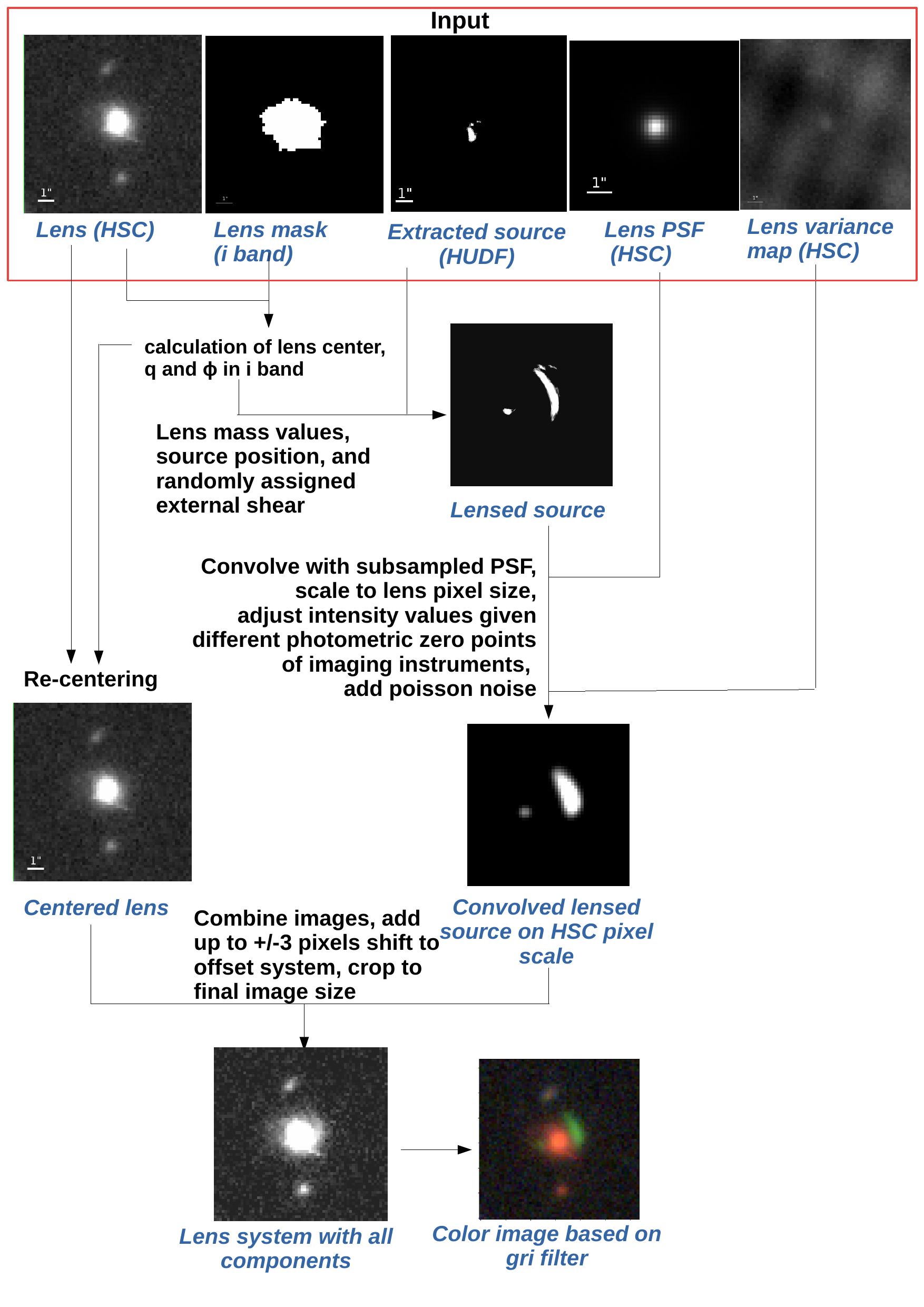}
    \end{center}
  \caption{Flow chart of the simulation pipeline used to create our training data in four filters $griz$. Figure adapted from Fig.~3 of \citetalias{schuldt21a}.\label{fig:flowchart}}
\end{figure*}

Training a supervised neural network requires a good-quality, representative, and sufficiently large sample of input data together with the corresponding output, which is the so-called ground truth. For our purpose, we request a sample containing  around 100,000 lens systems, each in four filters, which requires that we rely on simulations given that the number of real known lenses is $<10^4$ and the corresponding mass models need to be known. In order for our network to achieve optimum performance on real data, it is important that our mock images in our training sample are as realistic and as representative of real data as possible. We therefore use real observed images of galaxies as lenses and as background sources, instead of producing a completely mock sample as is often done \citep[e.g.,][]{hezaveh17, levasseur17, pearson19, pearson21}. We improve on the pipeline\footnote{The simulation pipeline will be made available upon request as well as the needed nonpublicly available software \GLEE \, \citep{suyu10a, suyu12b}.} described in \citetalias{schuldt21a} for simulating images of mock lenses and briefly summarize here the procedure, highlighting the new features we adopt in this work. A diagram of the pipeline is shown in Fig.~\ref{fig:flowchart}.

\begin{figure*}[ht!]
  \includegraphics[trim=0 0 0 0, clip, width=0.5\textwidth]{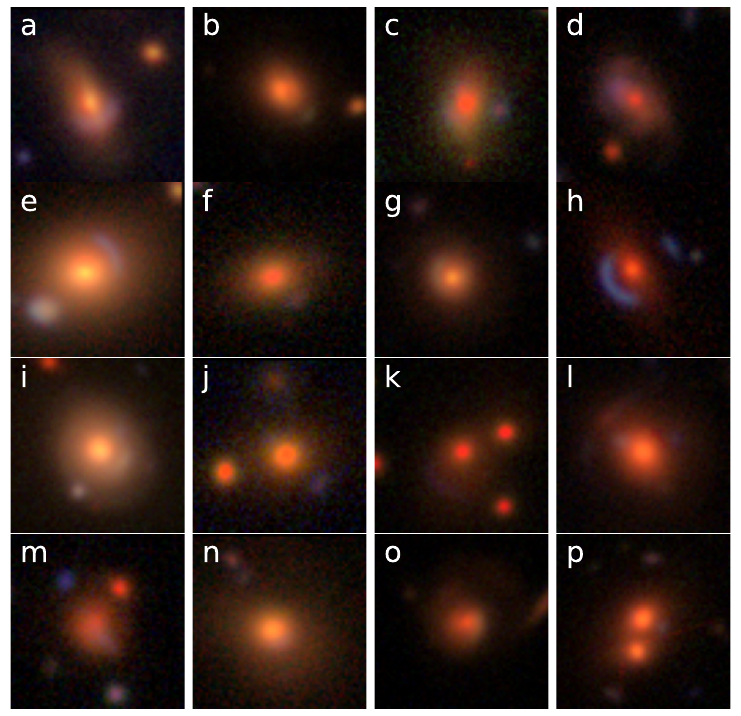}\includegraphics[trim=0 0 0 0, clip, width=0.5\textwidth]{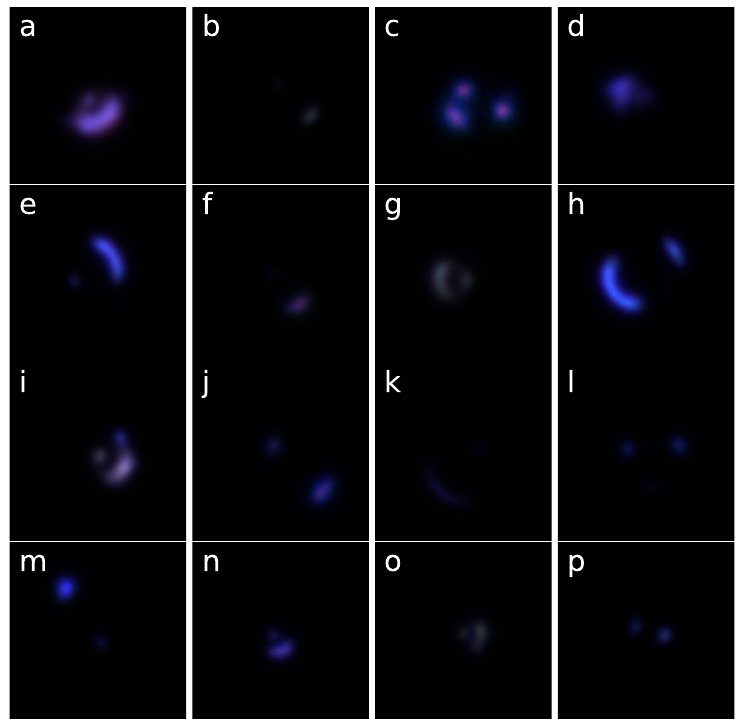}
  \caption{Example images of our mock lenses (left panel) and corresponding frames with only the lensed source (right panel). Each image has a size of $64 \times 64$ pixels, which corresponds to around $10.75 \arcsec \times 10.75\arcsec$. \label{fig:overview_newmocks}}
\end{figure*}

As lenses, we use again the HSC images \citep{aihara19} of luminous red galaxies (LRGs) and available spectroscopic redshifts and velocity dispersion measurements from the Sloan Digital Sky Survey DR14 \citep[SDSS,][]{SDSS14}. In addition to our criteria in \citetalias{schuldt21a}, we now set a lower limit of $100~\text{km} \, \text{s}^{-1}$ for the velocity dispersion, as very low-mass galaxies do not yield a strong lensing configuration with spatially resolved multiple images \citepalias[compare Fig.~1 in][]{schuldt21a}. This speeds up the selection process of suitable lens-source pairs. As background sources, we again use images from the Hubble Ultra-deep field \citep[HUDF,][]{beckwith06, inami17}. Each background galaxy is then lensed with the software \GLEE \, \citep{suyu10a, suyu12b}; convolved with the (subsampled) HSC PSF; and then binned to the HSC pixel size of 0.168\arcsec. The obtained arcs are then added to the lens image. With this procedure, we include real line-of-sight objects and light distributions, as shown in Fig.~\ref{fig:overview_newmocks}.

To calculate the deflection angles, we assume a singular isothermal ellipsoid (SIE) profile and, in contrast to \citetalias{schuldt21a}, an external shear described in complex notation by $\gamma_\text{ext,1}$ and $\gamma_\text{ext,2}$ to account for additional mass outside of the cutout. The convergence (also called dimensionless surface mass density) of the adopted SIE profile \citep{barkana98}\footnote{The SIE mass profile introduced by \citet{barkana98} allows for an additional core radius, which we set to $10^{-4}$ , yielding effectively a singular mass distribution without numerical issues at the lens center.} can be expressed as
\be
\kappa(r) = \frac{\theta_\text{E}}{(1+q)r}
,\ee
with an elliptical radius
\be
r = \sqrt{x^2 + \frac{y^2}{q^2}} \label{eq:ellipticalradius}
,\ee
and the axis ratio 
\be
q = \sqrt{ \frac{1-\sqrt{e_x^2 + e_y^2} }{ 1 + \sqrt{e_x^2 + e_y^2} } } \, ,
\label{eq:q}
\ee
where
\be
e_\text{x} = \frac{1-q^2}{1+q^2} \cos (2\phi)
\ee
and
\be
e_\text{y} = \frac{1-q^2}{1+q^2} \sin (2\phi)
\ee
are the complex ellipticities. The external shear parameters can be converted to a total shear strength
\be
\gamma_\text{ext} = \sqrt{ \gamma_\text{ext,1}^2 + \gamma_\text{ext,2}^2 }
\label{eq:gamext}
,\ee
which is rotated by
\be
\phi_\text{ext} = \left\{\begin{array}{cccc} \frac{s}{2} & \text{if } \gamma_\text{ext,1} \geq 0 \text{ and } \gamma_\text{ext,2} \geq 0\\ \frac{\pi - s}{2} & \text{if } \gamma_\text{ext,1} < 0 \text{ and } \gamma_\text{ext,2} \geq 0 \\ \frac{\pi+s}{2} & \text{if } \gamma_\text{ext,1} < 0 \text{ and } \gamma_\text{ext,2} < 0 \\ \frac{2\pi-s}{2} & \text{if } \gamma_\text{ext,1} \geq 0 \text{ and } \gamma_\text{ext,2} < 0 \end{array} \right. \
\label{eq:PAext}
\ee
with
\be
s = \arcsin \left( \frac{ |\gamma_\text{ext,2}|}{\gamma_\text{ext} } \right) .
\ee
Given our results in \citetalias{schuldt21a}, we again use a flat distribution of the Einstein radii up to $\simeq 2\arcsec$ in order to allow the network to better learn the full parameter space. In analogy, we use a flat distribution for $\gamma_\text{ext}$ in the range between 0 and 0.1, but consider also a realistic distribution \citep{faure11, wong10}.

Moreover, we improve our simulation code by including Poisson noise for the arcs approximated as
\be
\sigma_{\text{poisson,arc,}i} = \sqrt{\alpha \times I_{\text{arc,}i}^+}
,\ee
where $I_\text{arc}$ is the lensed source image (arcs) and
\be
I_\text{arc}^+ = \text{max}\{I_\text{arc},0\} \, .
\ee
To compute the scaling factor $\alpha$, we start from the HSC variance map with value $v_{i}^+$ corresponding to the $i^\text{th}$ lens image pixel\footnote{In the very rare case of negative values in the variance map $v_{i}$ provided by HSC, we reset them with zero, i.e., $v_{i}^+ = \text{max}\{v_{i},0\}.$} and define
\be
\sigma_{i} = \sqrt{v_{i}^+} \, .
\ee
We then compute the lens background noise $\sigma_\text{bkgr}$ defined as the minimal root-mean-square (rms) of the four corners of the lens image to exclude contributions from line-of-sight objects. With those two quantities, we then approximate the Poisson noise map of the lens as
\be
\sigma_{\text{poisson,}i} = \sqrt{\sigma_i^2 - \sigma_\text{bkgr}^2}
\ee
in order to obtain the Poisson scaling factor map,
\be
\alpha_i = \sigma_{\text{poisson,}i}^2/I_{\text{lens,}i}
,\ee
where $I_\text{lens}$ is the lens image from HSC. As this scaling factor $\alpha$ should be a constant, we approximate it as the median of the $10 \times 10$ pixel central region, because the lens intensity is highest in this region and therefore the mapping is more precise than on the outer parts. 
From this map, we draw Gaussian variations representing the additional Poisson noise, which are added on top of the simulated images.

Additionally, we improve the lens centering and ellipticity estimation in the simulation code. For this, the code now accepts a mask of the lens, obtained for example with the Source Extractor \citep{bertin96}, resulting in a more accurate masking of the lens. As that mask might exclude parts of the lens due to overlapping line-of-sight objects, we additionally apply a circular mask with a radius of 20 pixels centered at the image center when determining the lens center. From the resulting masked image, we predict the lens center through the first moments and re-center the image on the pixel closest to the lens center. The remaining fractional pixel offset of the lens center is taken into account through the simulation as lens \textit{light} center. The axis ratio $q_\text{ll}$ and position angle $\phi_\text{ll}$ of the lens \textit{light} is determined through the second moments using the provided mask.

As we now re-center the lens cutout, we shift the final mock image randomly by up to $\pm 3 $ pixels in the $x$ and $y$ direction such that the lens light and mass center, which are different, are not coincident with the image cutout center. This allows the network to learn the lens center instead of the cutout center and ensures that the network can handle images that are not perfectly centered on the lens mass.

Similar to \citetalias{schuldt21a}, we set some criteria on the brightness of the arcs to ensure visibility. Specifically, we request that the brightest pixel of the arcs is above 5$\sigma_\text{bkgr}$ in either $g$ or $i$ band, chosing the band in which the source is brightest. In the same filter, we further set a threshold for that pixel to be a factor of 1.5 brighter compared to the lens image at the same pixel. This helps to avoid drastic blending with the lens. To overcome these criteria more easily, we include a slight boost of the source magnitude in up to six steps of $-0.5$ magnitudes. This does not affect the shape of the arcs, nor the lens, nor the line-of-sight objects, but helps to obtain detectable lenses.

With this method, we generate around 165,000 mocks for training, validating, and testing the network, where we limit the Einstein radius to the range between $0.5\arcsec$ and $5\arcsec$, and additionally to 3,000 systems per 0.05\arcsec \, bin to obtain a flat distribution at least up to $\theta_\text{E} \sim 2\arcsec$. As we use real measurements of the velocity dispersion and redshifts, lensing systems with an Einstein radius above $\sim 2.5\arcsec$ are very rare, even when increasing the number of iterations for testing different lens--source alignments or lens--source pairs. As a result, the number of systems drops by two orders of magnitude towards $\theta_\text{E}\sim3\arcsec$ compared to the plateau. Example images generated with our upgraded simulation pipeline are displayed in Fig.~\ref{fig:overview_newmocks}; the left panels show the final mock images and the right panels show the lensed source alone, which were added to the HSC lens image. This figure demonstrates the variety of mocks and how realistic they are.

\FloatBarrier
\section{Neural networks and their architecture}
\label{sec:network}

% intro in ML but very brief

Convolutional neural networks are very powerful tools in image recognition tasks, especially if an autonomous and fast method is required to cope with huge numbers of images. This property has already lead to many different applications in astrophysics \citep[e.g.][]{paillassa20, tohill21, wu20, schuldt21b}. As there are already thousands of known lens candidates in HSC \citep[e.g.,][]{wong18, sonnenfeld19, sonnenfeld20, jaelani20a, jaelani20b, jaelani22, canameras21b, shu22}, and we expect hundreds of thousands more observed by LSST and Euclid \citep{collett15}, CNNs would be perfectly suited for analyzing this amount of data in an acceptable amount of time.

\begin{figure}[ht!]
\centering
  \includegraphics[trim=68 40 125 0, clip, width=\columnwidth]{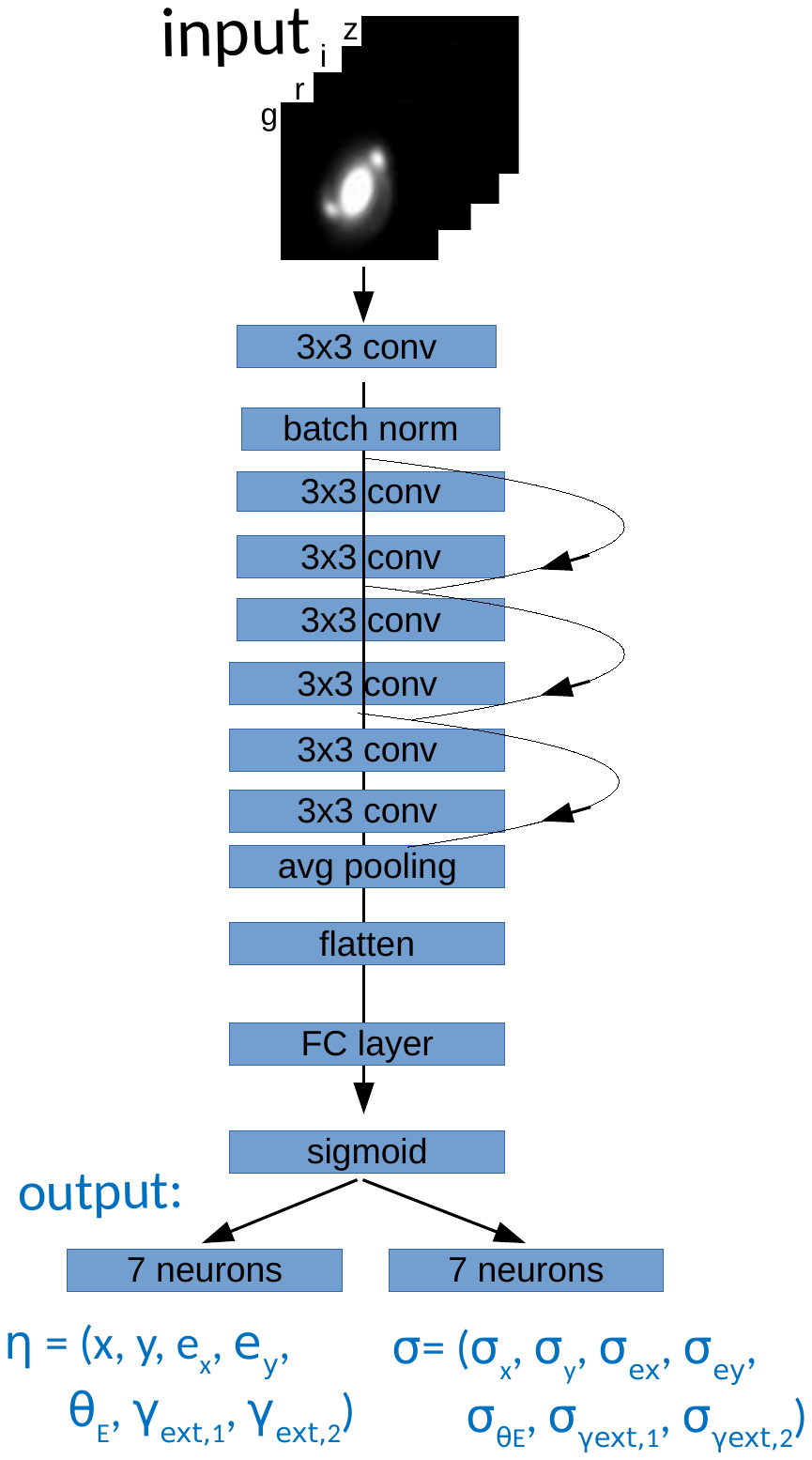}
  \caption{Overview of our ResNet architecture. The input are the lens images in four filters $g,r,i,$ and $z$, which are passed through a series of convolutional (conv.) layers and fully connected (FC) layers. After a sigmoid layer to map all output values to the range $[0,1]$, we split them up into median $\vec{\eta}$ and uncertainty $\vec{\sigma}$.\label{fig:networksketch}}
\end{figure}

While in \citetalias{schuldt21a} we used a CNN based on the LeNet \citep{lecun98} architecture to predict the five SIE parameters, deeper networks help to capture small features of the more complex mocks used in this work. Therefore, we now make use of residual neural networks \citep{he16a}, which are a specific type of CNN. In residual neural networks, the network architecture includes so-called residual blocks of typically two convolutional (conv.) layers with a kernel size of 3x3, which are connected via a skip-connection (or short cut) for the back-propagation. Therefore, the back-propagating gradient does not vanish easily over the multiple convolutional layers and thus allows  the neurons and weights of the first layers to be properly updated, even if the network architecture is very deep. A sketch of our network architecture is given in Fig.~\ref{fig:networksketch}. 

\subsection{Error estimation and loss function}
\label{sec:network:loss}

In analogy to \citetalias{schuldt21a}, we started with a network\footnote{The code used for the network training is based on a nonpublicly available code and will be made available upon request.} predicting one point estimate per parameter, which amounts to seven values for the five SIE parameters, plus two for the external shear parameters. Here we again used  a mean-square-error (MSE) loss function. We then modified the network such that it predicts two values per parameter, that is, 14 values in total. Providing not only a point estimate but rather a median-like value with a $1\sigma$ uncertainty makes the network much more powerful and helps to exclude insecure parameter estimations. From the 14 predicted values, we now interpret one value per parameter as the median (like previously the point estimate) and the other value as the standard deviation of a Gaussian function describing the error on that parameter value for that specific image. Here, we follow \citet{levasseur17} and use a loss $L$ given by
\be
L = \sum_{k=0}^N \sum_{l=0}^p \left[ -w_l \times P \left( \eta^\text{pred}_{k,l}, \eta^\text{tr}_{k,l}, \sigma_{k,l} \right)  + \epsilon_l \times \log \left( \sigma_{k,l}^2 \right) \right]
\label{eq:loss}
,\ee
with a regularization term to minimize the errors and a log-probability term that is defined in PyTorch for a Gaussian distribution, as %https://pytorch.org/docs/stable/_modules/torch/distributions/normal.html#Normal.log_prob
\be
P(\eta^\text{pred}_{k,l}, \eta^\text{tr}_{k,l}, \sigma_{k,l}) = - \frac{ \left( \eta_{k,l}^\text{tr} - \eta_{k,l}^\text{pred} \right) ^2}{2\sigma_{k,l}^2} - \ln(\sigma_{k,l}) -\ln(\sqrt{2\pi}) \, .
\ee
The loss for a given system is the sum over the $p$ different parameters $\vec{\eta} = (x,y,e_\text{x}, e_\text{y}, \theta_\text{E}, \gamma_1, \gamma_2)$ and corresponding errors $\vec{\sigma} = (\sigma_x, \sigma_y, \sigma_{e_\text{x}}, \sigma_{e_\text{y}}, \sigma_{\theta_\text{E}}, \sigma_{\gamma_1}, \sigma_{\gamma_2})$ with index $l \in [0,p]$. For the total loss $L$, we additionally sum over each image $k$ in the batch or sample of size $N$.

% test with w_rE = 5, (18196-18210)
%    -> not needed, rE already ok, shear the real problem.
    
% test to weight shear higher by factor 5, bot with rE factor 1 or 5 weighted
%  -> not really better on the shear, but loosing performance on other parameters
Similar to the weighting factors $w_l$ introduced already in \citetalias{schuldt21a} and used to improve on the Einstein radius, we have also introduced weighting factors in our new loss function as well as a regularization constant $\epsilon_l$ that can be different for each parameter. With these factors, we can control the contribution of the different parameters to the loss, and can therefore choose which ones to better optimize. For our CNNs in \citetalias{schuldt21a}, we found it helpful to increase the contribution of the Einstein radius to the loss, as this is the key quantity in the SIE profile. Although it remains the key parameter in an SIE+$\gamma_\text{ext}$ setup, we also tested the possibility of up-weighting the external shear, as these two parameters are the most problematic ones. However, we discarded this option from all further tests, as we find only a minor improvement on the external shear but notably lose performance on the other parameters. %As already discussed in Sect.~\ref{sec:results}, it seems that the external shear introduces too small distortions on the arcs that can be utilised by the network, making the external shear values difficult to recover.

%some tests without regularization, few tests with different regularization terms (18xxx)

%test different regularization terms, e.g. squared (normally and finally)
%    vs absolute value (20121-20150) -> loss not comparable to other ones
%    based on plot, good on SIE, but similar bad on shear 
    
% test different architectures: test with conv 1x1 (20181-20210)
%    best is 20183: very good on SIE, but completely flat on shear, loss -500,05, i.e. others are better but first time below -500! I would quantitatively from the plot say one of the best on SIE pararmeters
%    -> maybe worth trying further in this direction?
    
In addition to the weighting factors, we modified the loss function by introducing the uncertainty prediction such that also $\vec{\sigma}$ contributes to the optimization of the weights and neurons during the training process. Here, we tested different possible regularization terms, such as using the absolute value of $\vec{\sigma}$ instead of the squared term, or even leaving out the additional regularization term completely. Moreover, we varied the regularization constant $\epsilon$. As changing the loss function will change the loss value for a given network, we cannot compare the obtained loss values directly. Based on a more quantitative comparison, we found no notable difference in the performance on the median predictions $\vec{\eta}$. Given that we use a scaling to match the expected $68.3\%$ confidence intervals (CIs) instead of tuning dropout for this, we finally adopted the regularization function proposed by \citet{levasseur17}, which uses a squared term and a regularization constant of 0.5.

In order that the predicted $\sigma$ values can indeed be interpreted as $1\sigma$ Gaussian widths, \citet{levasseur17} suggest tuning the network through the dropout rate \citep{hinton12, srivastava14} such that the predicted errors match the expected CIs of 68.3\% for 1$\bar{\sigma}$, 95.4\% for 2$\bar{\sigma,}$ and 99.7\% for 3$\bar{\sigma}$. Here, the bar indicates that the standard deviation is computed from the medians of a full sample, for example, the test set, and does not correspond to the uncertainty $\vec{\sigma}$ predicted by the network for an individual lens system and parameter. This idea was also adopted by \citet{pearson21} for their CNNs. As there is only one dropout rate, both teams average over their three or four parameters when trying to match the expected percentile such that the individual errors still differ from the expected CIs (compare Fig.~2 of \citet{levasseur17} and Fig.~4 of \citet{pearson21}). This would be even more difficult for us with seven parameters. Instead of using dropout with the same rate for both convolutional layers and fully connected (FC) layers, we test only the effect of dropout for the FC layers as the dropout rate typically differs significantly between convolutional layers and FC layers \citep{hinton12, srivastava14}. For these reasons, we finally do not use dropout for the error scaling and instead incorporate a direct scaling of each parameter uncertainty such that the uncertainties match  the CIs on the test set as closely as possible.

\subsection{Network architecture}
\label{sec:network:architecture}

%architectures 101 and 104 with different strides: best 21144: SIE pars good recovered, shear as usual but one of the better runs, loss -524,14, net 104, strides 2,1,1
%- net 109 and 110, with different neuron_list values -> deeper network architectures than before 
%best: for 109: 21254, loss -505,92 with neuron list #3, in general good on SIE, but very bad on shear
%best for 110: 21264, loss -506,08 with neuron list #5, good on SIE, similar good on shear as usual, but not bad

%Neuron option 1: 8,25,32,48,64,32,16
%option 2: 16,32,48,64,128,64,32
%option 3: 16,25,48,32,64,32,16
%option 4: 8,16,24,32,48,32,64,32,16
%option 5: 8,16,24,32,48,64,96,64,32
%option 6: 16,32,48,64,128,512,1024,128,32

In general, to find the best network architecture and the best set of hyper-parameters, which is defined as the network with minimal mean validation loss, we carried out extensive tests including the hyper-parameter search discussed in Sect.~\ref{sec:network:hyperparameter}. For the architecture, we varied the number of residual blocks between 2 and 6, the number of FC layers between 1 and 3, and the number of feature maps and strides in the convolutional layers and the number of neurons within the different FC layers. We also tested different kernel sizes for the convolutional layers, but obtained the smallest average validation loss with the standard $3\times3$ kernel, which is understandable given the small size of our ground-based images. 

As shown in Fig.~\ref{fig:networksketch}, the final network architecture contains one convolutional layer with kernel size 3 followed by a batch normalization and three residual blocks of two convolutional layers each with a kernel size of 3. As shown in Fig.~\ref{fig:networksketch}, we finally include three residual blocks with strides of two, one, and one, respectively, and 24, 32, and 64 feature maps, respectively. The first layer before the residual blocks has 16 feature maps, while the input has 4 feature maps corresponding to the four different filters. After flattening, the output of the convolutional sequence is passed through one FC layer connecting 1024 to 14 neurons. 

% test different poolings: max vs avg vs no at all, assume always 8x8 after all conv layer (20211-20240)
%    -> avg definitefly the best, but maybe smaller kernel better for max pooling
% also with samller pooling size, max pooling not really working...

During our tests on the network architecture, we also adjusted the pooling layer before flattening for the FC layers (compare Fig.~\ref{fig:networksketch}). Here, we tried an average pooling layer, a maximal pooling layer, or no pooling at all. For the two different pooling layers, we tested kernel sizes of $8 \times 8$, $4\times 4$, or $2\times2$. We found the best performance using an average pooling layer with a kernel size of $8 \times 8$, as indicated in Fig.~\ref{fig:networksketch}.

\subsection{Hyper-parameter search}
\label{sec:network:hyperparameter}

% range of lr values typically tested
% test decreasing lr, each 20 iterations factor 0.5

In addition to testing different architecture layouts, the values for hyper-parameters also need to be optimized. Throughout these tests, we adopted a weight decay of 0.0005, a momentum of 0.9, and, after some short tests regarding the effect on the performance, a batch size of 32. Apart from those, the learning rate is one of the key quantities to test. Here, we typically tested the values for the learning rate $r_\text{learn} \in [0.01, 0.001, 0.0001, 0.00001, 0.000001]$ when changing any other hyper-parameter or layer. This covers a good range of plausible values. As the changes on the weights are expected to drop over the training, we also tested the option of a ``decreasing learning rate'', which means that we divide the learning rate by a factor of two every twentieth epoch. Because the best epoch was typically below 100, we had only a few relevant decreasing steps during training such that we finally adopted a constant learning rate for our further tests. In the presented network, we finally assumed a learning rate of $r_\text{learn}=0.000001$, a regularization constant $\epsilon$ of 0.5 for all the different parameters, and a weighting factor $w$ of 5 for the Einstein radius in analogy to \citetalias{schuldt21a}, and otherwise 1.

%- test different seeds of a given setup (19061-19090)
%    setup: batch 32, no dropout, sigmoid normalization, sqrt-squeezing, reg const of 0.5, subsampling factor 2, up-weighting of shear by factor 5
%    -> each network obtains different loss values, but from performance plot very similar
%    -> additionally we see that upweighting of shear does not help for the shear
%    
%    - also (20031-20060), assume net 101 and (20061-20090) assumed net 104
%    -> loss always slightly different, but similar, also best epoch changes, but similar
%    -> network find not the same weights!
    
Although the initialization is not a hyper-parameter that typically gets tuned, we tested the effect of using a different initialization of the network for a few given setups. This demonstrates how independent the final, trained network is from the original values. We find no preference of a specific seed and the overall performance is unaffected, but each network gives a slightly different loss. These slight changes are due to the stochastic learning process and are therefore commonly observed. For a few instances, the best hyper-parameter, for example the learning rate, changed by changing the seed, indicating the importance of optimizing the hyper-parameters for a given network.

To mitigate these changes, so-called ensemble learning methods can be used, where essentially the same network, that is, with fixed architecture and hyper-parameters, is trained with a different initialization and the predictions are combined afterwards. We performed such tests for a few given setups, predicted the 14 parameters with each network, and compared their average to the ground truth on the test set. In our case, this does not help to eliminate outliers and decrease the scatter, which proves the similarity of the networks regardless of the initialization.

\subsection{Parameter normalization}
\label{sec:network:normalization}

Because different parameters cover different ranges, we include a scaling to map them consistently to the range $[0,1]$. This ensures an equal contribution from the different parameters to the loss, resulting in a better optimization of all parameters. We assume the following input ranges $[a_l,b_l]$: lens center $x \in [-0.6\arcsec,0.6\arcsec]$ and $y \in [-0.6\arcsec,0.6\arcsec]$, complex ellipticity $e_\text{x} \in [-1,1]$ and $e_\text{y} \in [-1,1]$, Einstein radius $\theta_\text{E} \in [0.5\arcsec, 5\arcsec]$ as already in the simulation procedure, and complex external shear $\gamma_\text{ext,1} \in [-0.1, 0.1]$ and $\gamma_\text{ext,2} \in [-0.1, 0.1]$. This means the ground truth is scaled  componentwise as
\be
\vec{\eta}^\text{scaled, tr} = \frac{\vec{\eta}^\text{tr} -\vec{a}}{\vec{b}-\vec{a}}\, ,
\ee
and the output of the network is scaled back to the original ranges through
\be
\vec{\eta} = (\vec{b}-\vec{a}) \, \vec{\eta}^\text{scaled, pred} +\vec{a,}
\ee
and for the uncertainty,
\be
\vec{\sigma} = (\vec{b}-\vec{a}) \, \vec{\sigma}^\text{scaled} \, .
\ee
The uncertainties are not shifted by $\vec{a,}$ as those are considered with respect to the predicted median values $\vec{\eta}$. To ensure that all predicted values are between 0 and 1, we include a sigmoid layer in the network architecture before splitting it up into seven values for the median $\vec{\eta}$ and seven values for the uncertainty $\vec{\sigma}$. The network parameter optimization is performed with a ReLU \citep{nair10} activation function and a stochastic gradient descent algorithm, adjusting the weights to minimize the loss.

%overall best: 21144 with loss -524,14, net 104, strides 2,1,1, epoch 49,
%-batch 32 -v -lr 0.00001 -net 104 -subfactor 1 -w 1,1,1,1,1,1,1 -norm sigmoid -s 1 -neuron_list 16,24,32,64,32,16 -regconst 0.5 -layer 2,2,2 -stride 2,1,1 -e 501 

%net 104: 1 conv layer + 3 conv layer blocks, avg pooling 8x8, 3 Fc layers, sigmoid scaling

\subsection{Cross-validation}
\label{sec:network:cross-validation}

To avoid unbalanced optimization of the network, we follow \citetalias{schuldt21a} and use a five-fold cross-validation by splitting our set of 165,374 mocks into roughly 56\% training, 14\% validation, and 30\% testing, where rounding effects from the batch size occur, and train each run over 500 epochs. This also allows us to better determine the hyper-parameters (e.g., learning rate, number of neurons, or feature maps) and the final stopping epoch, which is defined through the minimal average validation loss.

% test dropout, no much effect, refrain from using dropout for now, want to use it for error scaling (19001-19030)
%    also (20001-20030), fail on ext shear completely

The lowest mean loss is $-593.71$, obtained in epoch 230, such that the final run is trained over 230 epochs. The loss curve reveals that there is a relatively large generalization gap, which means that  after a certain number of epochs the network performs significantly better on the training data than on the validation set. Therefore, we tested the effect of dropout on the FC layers even with our relatively small number of FC layers. This helped to reduce the generalization gap, but also resulted in a higher average validation loss used to select the best network. Therefore, we consider no dropout for the final network, which was suggested by \cite{levasseur17} and also used by \citet{pearson21} for the uncertainty scaling such that the distribution matches the expected percentile for a Gaussian distribution.

\FloatBarrier
\section{Network results and performance}
\label{sec:results}

\begin{figure*}[ht!]
\begin{center}
\includegraphics[trim=0 440 0 0, clip, width=0.5\textwidth]{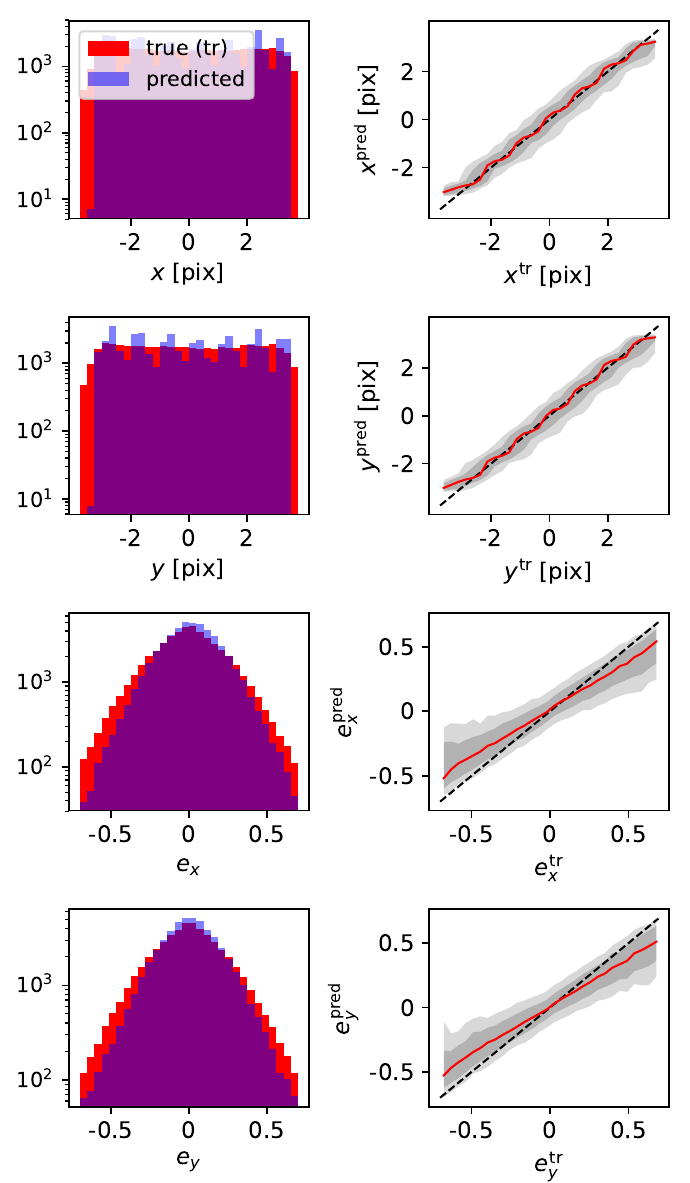}\includegraphics[trim=0 14 0 570, clip, width=0.5\textwidth]{plots/Network_22165/out_cornerplot_22165_final.pdf}
\caption{Comparison between the ground truth and prediction of our final network. In the left panel of each column we show histograms of the ground truth (red) and predicted values (blue). In the right panel we directly plot the predicted value ($y$-axis) against the true value, showing the median as a red line with 1$\bar{\sigma}$ and 2$\bar{\sigma}$ ranges (gray shaded) inferred from the distribution of median values of the test set.\label{fig:comparison} }
\end{center}
\end{figure*}

In this section, we present the predicted SIE+$\gamma_\text{ext}$ parameters and corresponding uncertainties
$\vec{\sigma}$ predicted with our residual neural network. This network was trained, validated, and tested on 165,374 realistic mock images created with our upgraded simulation procedure as described in Sect.~\ref{sec:simulation}. The optimized network architecture is shown in Fig.~\ref{fig:networksketch}. In Fig.~\ref{fig:comparison} we show a comparison between ground truth and predicted values on the test set, which are images that the network has not seen before during training or in the cross-validation procedure. Specifically, for every parameter, we show  the distribution of values as a histogram as well as a direct comparison between predicted and true values. Figure~\ref{fig:comparison} also shows the median predicted values per bin (red line) and the $1\bar{\sigma}$ and $2\bar{\sigma}$ (gray shaded). The bar indicates the computation from the whole sample to better distinguish from the specific uncertainties $\vec{\sigma}$ predicted by the network.

The network performs very well on the lens center, especially in contrast to the CNNs presented in \citetalias{schuldt21a} which had difficulty in recovering the mass center. Here, we clearly see the improvement on the lens centering resulting from the re-centering of the lens galaxy, taking the fractional pixel in the simulation into account, which was neglected in \citetalias{schuldt21a}, and the random shift of the final mock image by up to $\pm 3$ pixels. The median with 1$\bar{\sigma}$ range for $x^\text{pred} - x^\text{tr}$ and $y^\text{pred} -y^\text{tr}$ are, respectively, $  0.04 ^{+ 0.50 }_{ -0.40 }$ and $  0.04 ^{+ 0.46 }_{ -0.41 }$ pixels. The complex ellipticity components $e_\text{x}$ and $e_\text{y}$ are also well recovered with $ 0.02 ^{+ 0.12 }_{ -0.09 }$ and $  0.01 ^{+ 0.11 }_{ -0.10 }$, respectively, although the predictions still have a tendency to underestimate the absolute values. In other words, the network predicts the galaxies to be rounder than they are. One of the reasons for this is the much higher number of training systems with values around zero, which means that the network might be biased towards this value. 

\begin{figure*}[ht!]
\begin{center}
\includegraphics[trim=0 0 0 0, clip, width=0.8\textwidth]{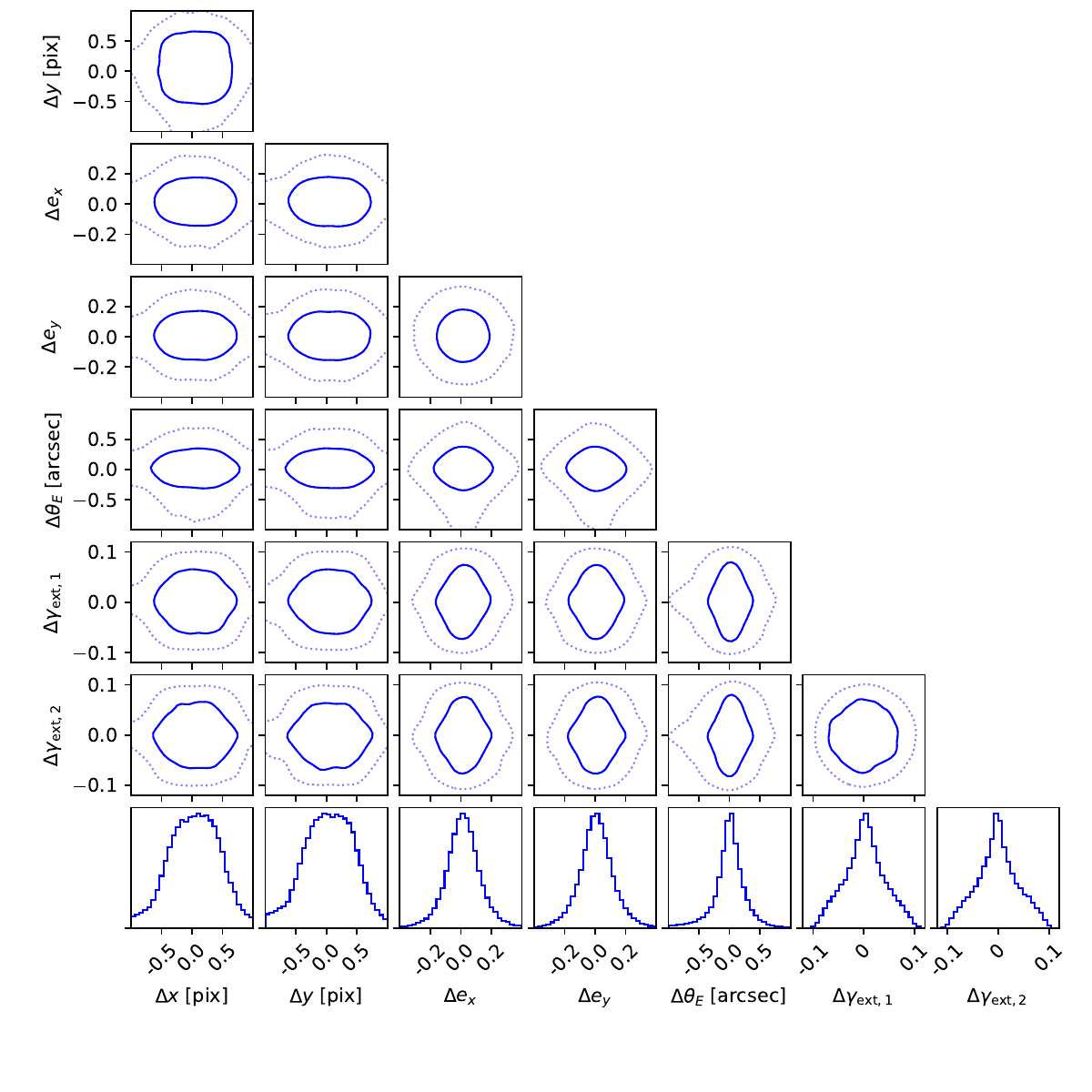}
\caption{Histograms (bottom row) and 2D density plots of the difference between prediction and ground truth of our final ResNet applied to the test set. \label{fig:cornerplot} }
\end{center}
\end{figure*}

The Einstein radius $\theta_\text{E}$ is very well recovered with a median and 1$\bar{\sigma}$ value of $ \theta_\text{E}^\text{pred}- \theta_\text{E}^\text{tr} = 0.003 ^{+ 0.21}_{ -0.24}$. This can also be seen from Fig.~\ref{fig:comparison}, where the median line closely follows the 1:1 line between our lower limit of $0.5\arcsec$ and around $1.5\arcsec$, dropping for systems with very large image separations. 
The 1$\bar{\sigma}$ and 2$\bar{\sigma}$ ranges indicate very precise predictions between $\sim1\arcsec$ and $\sim2\arcsec$, while beyond this range the performance is not as good, which is partly due to the low number of systems in the training set. The lower precision on the low end is most likely due to blending issues with the lens. Because of the small image separations, the counter images are more strongly blended with the lenses such that the network cannot sufficiently deblend the arcs from the lens, which is important for the prediction of the Einstein radius. This is confirmed by a test using the arcs alone for the network training (see Sect.~\ref{sec:tests:inputdata} for details). On the other side of the range (i.e., high $\theta_\text{E}^\text{tr}$), the performance drops significantly due to their underrepresentation in the data set. As shown with CNNs in \citetalias{schuldt21a}, the distribution of the Einstein radii is crucial for the performance, and a uniformly distributed sample yields the best performance over the full considered range. As mentioned already in Sect.~\ref{sec:simulation}, we therefore set a maximum of mocks per Einstein radius bin. As we use real measurements of the velocity dispersion and redshifts, lensing systems with an Einstein radius above $\sim 2\arcsec$ are very rare, such that the number of systems drops by more than one order of magnitude towards $\theta_\text{E}\sim3\arcsec$ compared to the plateau at $\theta_\text{E} \leq 1.5\arcsec$, which means a decreasing performance. Given the very low number of systems within each bin for $\theta_\text{E} > 3\arcsec$, % and because of the increasing chance of group-scale lenses for which systems the network is not trained,
we do not show them in Fig.~\ref{fig:comparison} but keep our limit at $5\arcsec$ as the largest image separation accepted by our simulation pipeline. Even if the network shows a lower performance in this range, it has seen some of these systems, and because of our introduced scaling, it is in principle able to predict such large Einstein radii. 
As demonstrated in \citetalias{schuldt21a}, we can also train a dedicated network on a smaller sample, for example for systems with $\theta_\text{E} \geq 2\arcsec$, meaning that a better performance is achieved for systems with such large image separations.

As we see from Fig.~\ref{fig:comparison}, the network is so far not able to accurately predict the external shear components $\gamma_\text{ext,1}$ and $\gamma_\text{ext,2}$, although the mean with 1$\bar{\sigma}$ values for the whole test set are $  0.002 ^{+ 0.04}_{ -0.04}$ and $  -0.001 ^{+ 0.04}_{ -0.04}$, respectively. It tends to predict values closer to zero, resulting in a lower predicted shear strength than it should. We tested many different possibilities to improve on the external shear as described in Sect.~\ref{sec:tests}, and found that blending with the lens is not the main reason for this issue. It seems that the current network apparently cannot sufficiently generalize to new systems on these very minor distortions on the arcs. This might be because of the variable PSF from system to system or the image resolution given that it works relatively well with more idealized, lens-light subtracted and high-resolution images \citep{morningstar18}. Further investigation beyond our tests summarized in Sect.~\ref{sec:tests} is therefore necessary for a better estimate of the external shear. Whether a precise estimate of the external shear is crucial depends on the science case behind the modeling. For statistical studies on the lenses, for instance regarding the lens mass, the external shear is expected to have only negligible influence.

%median +/- 1 sigma  $  -0.05631531987871469 ^{+ 0.4057232822690689 }_{ -0.5393879754202706 }$
%median +/- 1 sigma  $  0.1428352934973579 ^{+ 0.618688975061689 }_{ -0.35395792552403105 }$

%median +/- 1 sigma  $  -0.00856626033782959 ^{+ 0.09951460361480713 }_{ -0.11409062147140503 }$
%median +/- 1 sigma  $  0.024280905723571777 ^{+ 0.12822628021240234 }_{ -0.08380591869354248 }$

%median +/- 1 sigma  $  0.0015576928853988647 ^{+ 0.09042677097022533 }_{ -0.08893319964408875 }$

%median +/- 1 sigma  $  0.0012932479381561335 ^{+ 0.03599641919136048 }_{ -0.03367191553115845 }$
%median +/- 1 sigma  $  -0.001070988178253171 ^{+ 0.03398900926113129 }_{ -0.03574174642562866 }$

Figure~\ref{fig:cornerplot} shows the difference between the predicted values and ground truths for the seven parameters, as well as correlations between them. We find no strong correlations, not even between typically degenerate quantities such as ellipticity and external shear. By comparing to Fig.~7 of \citetalias{schuldt21a}, where the plotting ranges of the SIE parameters are kept the same for ease of comparison, we find a generally better performance on the Einstein radius but with the same kind of diamond-shaped 2$\bar{\sigma}$ contour. On the other hand, the scatter on the lens mass center is slightly larger with the presented ResNet. This is  most likely because of our newly introduced random $\pm 3$ pixel shift for the final mocks, which was implemented to ensure that the network predicts the lens mass center instead of the image center, but means that we cannot directly compare the performance on the lens center here. Instead, Fig.~\ref{fig:comparison} reveals the much improved performance on the lens center of the ResNet compared to the CNNs in \citetalias{schuldt21a} (Fig.~8 and Fig.~10). Given the numerous changes between these networks by introducing new parameters through the external shear, the uncertainty prediction, the change in the network architecture, and the addition of Poisson noise in the simulations, it is difficult to interpret the differences in performance. As the ResNet has the much more complex task of additionally predicting the external shear and uncertainties for each parameter, which it does
well overall, this is definitely the more powerful network.

%% PLOT SHOULD BE HERE

Besides the median value, the network also predicts an uncertainty $\sigma$ for each parameter. To interpret this as the width of a Gaussian distribution, it has to match the statistical expectations of 1$\bar{\sigma}$ corresponding to a CI of $68.3\%$, 2$\bar{\sigma}$ to $95.4\%$, and 3$\bar{\sigma}$ to $99.7\%$. As explained in Sect.~\ref{sec:network}, we do not use dropout, as in \citet{levasseur17} and \citet{pearson21}, and instead incorporate a direct scaling per parameter. As our predictions do not perfectly match a Gaussian distribution, we scale the predicted values for the different parameters $\vec{\sigma}$ = ($\sigma_x$, $\sigma_y$, $\sigma_{e_\text{x}}$, $\sigma_{e_\text{y}}$, $\sigma_{\theta_\text{E}}$, $\sigma_{\gamma_\text{ext,1}}$, $\sigma_{\gamma_\text{ext,2}}$) by $\vec{s} = (1.25, 1.32, 1.19, 1.20, 1.08, 1.21, 1.21)$. This minimizes the quadratic sum of the differences for the three $\sigma$ intervals for each parameter $\eta_j$, that is, it minimizes
  \be
  d_{1,j}^2+d_{2,j}^2+d_{3,j}^2
  ,\ee
  with
  \bea
  d_{1,j} = \left| \frac{100}{N} \times \sum_k^T u_{1,j,k} - 68.3\right|\\
  d_{2,j} = \left| \frac{100}{N} \times \sum_k^T u_{2,j,k} - 95.4\right|\\
  d_{3,j} = \left| \frac{100}{N} \times \sum_k^T u_{3,j,k} - 99.7\right|
  ,\eea
  where $T$ denotes the size of the test set over which we sum, and
  \bea
  {u}_{1,j,k} =&\left\{\begin{array}{ccc} 1 &  \text{if} & \left( |{\eta}_{j,k}^\text{tr} - {\eta}_{j,k}^\text{pred}| - s {\sigma}_{j,k} \right) < 0\\
  0 & \text{otherwise} & \end{array} \right. \\
  {u}_{2,j,k} =&\left\{\begin{array}{ccc} 1 &  \text{if} & \left( |{\eta}_{j,k}^\text{tr} - {\eta}_{j,k}^\text{pred}| - 2s {\sigma}_{j,k} \right) < 0\\
  0 & \text{otherwise} & \end{array} \right. \\
  {u}_{3,j,k} =&\left\{\begin{array}{ccc} 1 &  \text{if} & \left( |{\eta}_{j,k}^\text{tr} - {\eta}_{j,k}^\text{pred}| - 3s {\sigma}_{j,k} \right) < 0\\
  0 & \text{otherwise.} & \end{array} \right.
  \eea

\begin{figure}[ht!]
\begin{center}
  \includegraphics[trim=0 0 0 0, clip, width=0.5\textwidth]{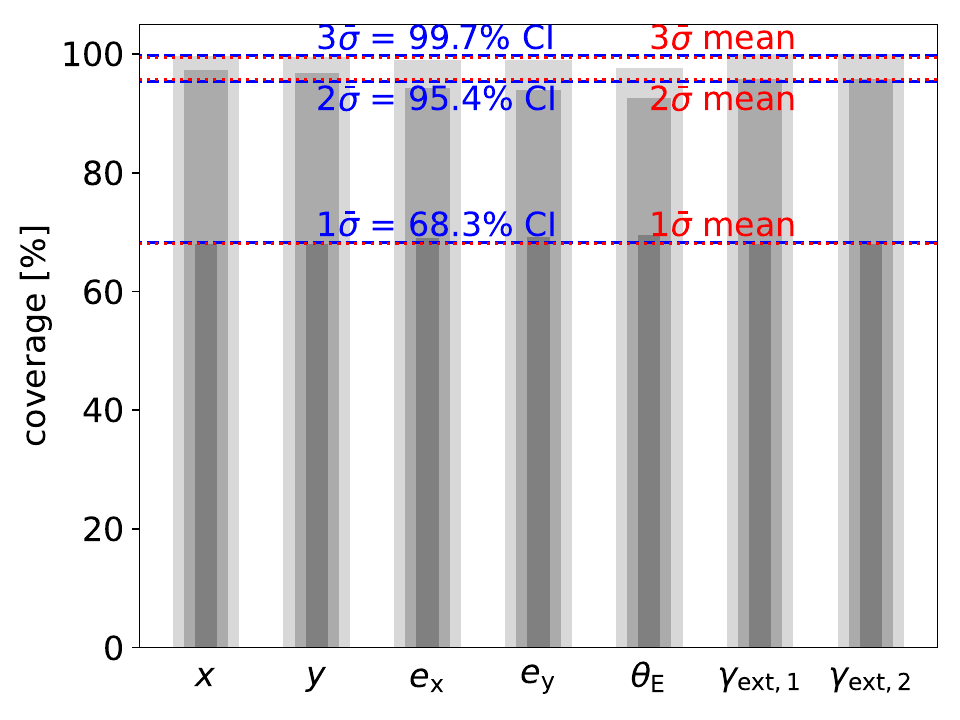}
  \includegraphics[trim=0 0 0 0, clip, width=0.5\textwidth]{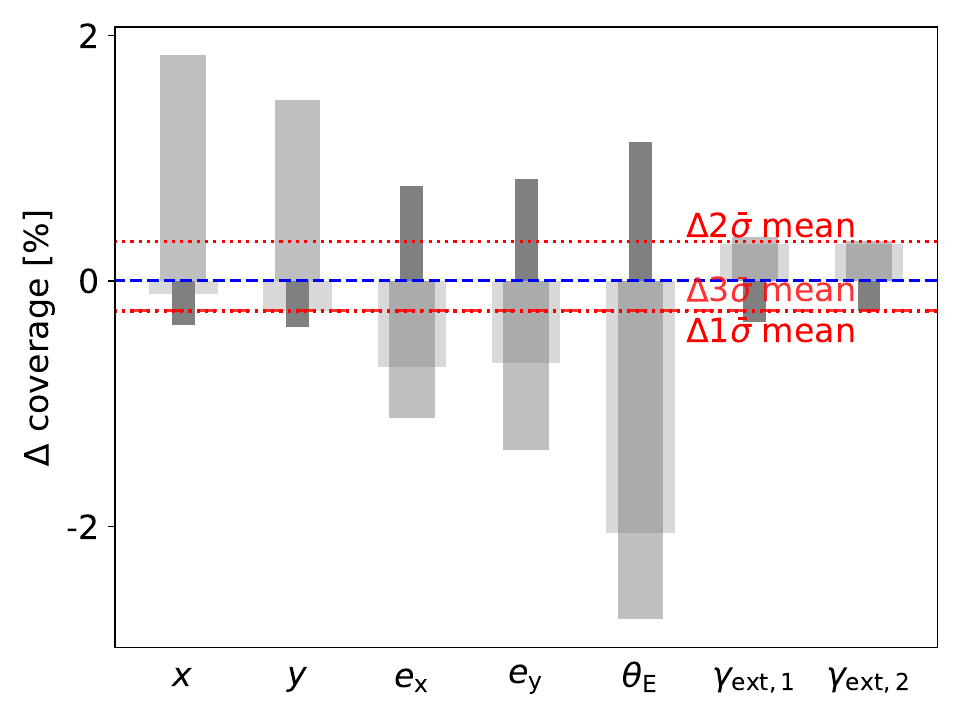}
  \caption{Coverage of the predicted uncertainties, shown as absolute values (top) and relative values (bottom) to expectations (blue dashed lines). The network-predicted uncertainties $\vec{\sigma}$ of the test set were scaled such that roughly 68.3\%, 95.4\%, and 99.7\% of the true values $\vec{\eta}^\text{tr}$ are contained in the 1$\sigma$ (dark gray bar), 2$\sigma$ (median gray bar), and 3$\sigma$ width (light gray bar) ranges of a Gaussian distribution centered at the predicted median values $\vec{\eta}^\text{pred}$. With this scaling, we can indeed interpret the predicted uncertainties of each individual parameter as the width of a Gaussian distribution. We additionally show the mean of all seven $\sigma$ values, which are, respectively, 68.05\%, 95.72\%, and 99.46\%, in red for comparison.\label{fig:coverage}}
\end{center}
\end{figure}

  This is a good compromise between matching the commonly used 68.3\% CIs and also the 95.4\% and 99.7\% CIs. This is visualized in Fig.~\ref{fig:coverage}, where we show the coverage of the scaled uncertainty values for each parameter (gray bars) as absolute values (top) and the difference from the expectations (bottom). The top panel demonstrates the close match between the scaled uncertainties and the expected CI levels (blue dashed), especially for the mean over all seven parameters (red dotted), and can be directly compared to the achievements from \citet[][Fig.~2]{levasseur17} and \citet[][Fig.~4]{pearson21}. The bottom panel highlights the small deviations between the achieved and expected CIs; it shows the good match of the 1$\sigma$ values for all individual parameters achieved through appropriate scaling, resulting in visible deviations for the 2$\sigma$ and 3$\sigma$ lines. In particular, the distribution for the Einstein radius is sharper than a Gaussian distribution.

A comparison of the performance to other modeling networks is difficult given the discrepancy in assumptions. \citet{hezaveh17}, who originally proposed this novel idea, presented a network to predict $e_\text{x}$, $e_\text{y}$, and $\theta_\text{E}$ of an SIE profile for HST-like images after lens-light subtraction to demonstrate the feasibility. \citet{levasseur17} further included uncertainty predictions and an external shear component. Similarly, \citet{pearson21} presented a network to predict $e_\text{x}$, $e_\text{y}$, and $\theta_\text{E}$ of an SIE profile for mock images with $0.1\arcsec$ resolution in preparation for the Euclid space mission. These authors also included error estimations inspired by \citet{levasseur17} and explored the opportunity of a hybrid code by combining their network with \Pyautolens \, \citep{nightingale18}, a fully automated nonmachine-learning-based modeling software, for further refinement of the parameter predictions. The difference in image resolution, number of filters, and the quality of the training and test data makes a comparison difficult. Moreover, the different number of predicted parameters complicates the comparison given the degeneracies between the different parameters of a given lensing system. The closest work in terms of image quality and number of filters was presented by \citet{pearson19}, who considered CNNs to predict $e_\text{x}$, $e_\text{y}$, and $\theta_{E}$ of an SIE profile for Euclid, LSST $r$-band, and LSST $gri$-band data. The latter is the best match to our networks, and is comparable in performance, as mentioned in \citetalias{schuldt21a}. There is also currently a lot of work going into automated modeling without machine learning \citep[e.g.,][]{nightingale18, nightingale21b, rojas22, savary22, ertl22, etherington22, gu22, schmidt23}, which typically performs better than neural networks but requires significantly longer run times of hours to days. We refer to \citet{schuldt22b} for a direct comparison between the network presented here and traditionally obtained models for real HSC lenses.

\FloatBarrier
\section{Additional tests on the network}
\label{sec:tests}

In this section, we summarize additional tests carried out mostly to address the remaining difficulty in the prediction of the external shear.

\subsection{Tests on the network architecture}
\label{sec:tests:architecture}

% paralell FC: best 21319, loss -508,18, net 120, SIE ok, shear ok but very flat (i.e. not that good)
%net 120: # neuron list e.g., 8,16,24,32,48,32,64,32,16; layer 2,2,2,2,2,2; stride 2,2,2,2,2,2; (6conv+3FC) - similar as 104 but three conv layer more, 2 paralell FC layers
%test different versions of paralell FC layers: 7 branches, 14 branches, 2 branches
%
%net with 7 branches: 121, 118, 115, 112, -> all not good
%net with 14 branches: 122, 119, 116, 113 -> all not good
%net with 2 branches: 120, 117, 114, 111 -> always much better than 7/14 branches

\begin{figure*}[ht!]
\begin{center}
\includegraphics[trim=0 150 0 80, clip, width=\textwidth]{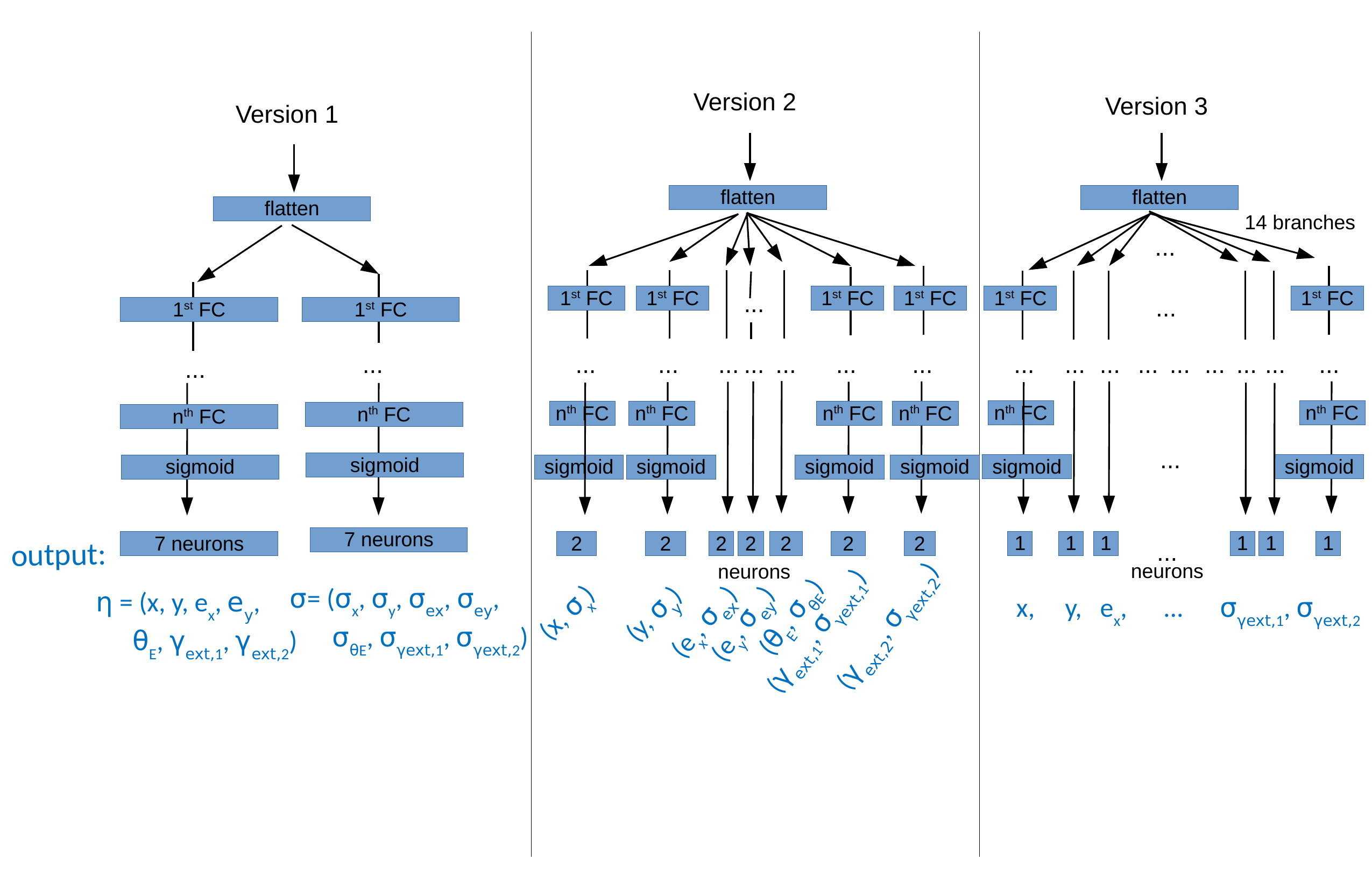}
\caption{Sketch of tested multibranch network architectures. We split the FC layers either into 2, 7, or 14 branches after the convolutional layers and flattening (compare Fig.~\ref{fig:networksketch}).\label{fig:multibranch} }
\end{center}
\end{figure*}

Beyond our general tests on the network architecture described in Sect.~\ref{sec:network}, we further tested the possibility to split the network into multiple branches after flattening and before the FC layers as sketched in Fig.~\ref{fig:multibranch}. Each branch then consists of $n$ FC layers, which can also mean just one, and predicts only specific parameters. This allows us to optimize the weights of the different branches for the specific parameters of that branch. The input of the first FC layer in each branch is the full flattened data cube obtained after the pooling, which is the same for each branch. Here we considered three versions visualized in Fig.~\ref{fig:multibranch}, although others are possible. In version 1, we split the network into two branches, each predicting seven values: one branch for the median values $\vec{\eta}$ and one branch predicting the uncertainties $\vec{\sigma}$. In version 2, we split the network into seven branches, each predicting the median and uncertainty for one parameter. The third considered option includes 14 branches, where each branch predicts just one value. As we did not find an improvement through these tests, we do not further explore the splitting into multiple branches. This might be because the parameters have shared information and are degenerate, such that a single branch is helpful especially for the uncertainty prediction. However, even though the multiple branches were not helpful in obtaining improved performance compared to an architecture with just a single branch of FC layers, they could be helpful when trying to tune the errors through the dropout rate as suggested by \citet{levasseur17}, as one can adopt a specific dropout rate for each branch. This would allow an individual tuning for each parameter.

\subsection{Over-fitting tests}
\label{sec:tests:overfitting}

% overfitting tests:
% - test different architectures 
% - test for shear
%   -> if everything else fixed, can recover shear
%   -> as soon as other parameters vary, network fails on the shear
%   
%   - with 9649 images (19121-19150) plus shear up-weighting by factor 5
%    -> ok on center and ex and ey, quite bad on shear, but also on rE -> can't learn to find the arcs
    
To test whether a particular network architecture is promising, we performed so-called over-fitting tests, which means that we trained and evaluated the network on a very small sample with around 1,000 mock images. This shows whether the network is able to ``memorize'' the task perfectly, including predicting the external shear. As these were only short tests, we performed no cross-validation. Given our difficulties with the external shear, it is important to remark that our network is indeed able to learn the external shear for that very small sample. This shows that the network is in general able to extract features of the images and to connect them to all seven parameters, albeit imperfectly. On the shear in particular we see some scatter, which indicates that the network does not just remember the exact images and output the stored values. This demonstrates that the baseline network architecture is likely not the main reason for the failure in the external shear prediction. However, these tests have no implications as to whether the network performs well on new, unseen data from the test set. We further performed such over-fitting tests with networks that just predict the external shear. This helped us to significantly improve on the training data. 
    
\subsection{Test with fixed lens--source pairs}
\label{sec:tests:fixpairs}

As our over-fitting tests presented in Sect.~\ref{sec:tests:overfitting} only show that the network is able to predict the external shear well from images in a small training set, we further tested whether it can predict the shear on new images if we simplify the task. For this, we considered three different stages of simplification and created samples with 1,000 mocks each. First, we always used the same lens, but different background sources from HUDF and placed them randomly behind the lens. The second scenario had the same lens light and source light distribution, as well as the same redshifts, but various positions of the background source were tested with respect to the lens as well as different mass-to-light offsets of the lens, as in our general training data set. The third option was to keep everything fixed, including the source position, and only vary the external shear. This means that, in the third option, the arcs always appeared to be the same, without external shear, and only distortions arose from the external shear. As the lens did not vary in these tests, we excluded the SIE parameters from the prediction and trained the networks to only predict the external shear.

\begin{figure}[ht!]
\begin{centering}
    \includegraphics[trim=0 400 0 0, clip, width=\columnwidth]{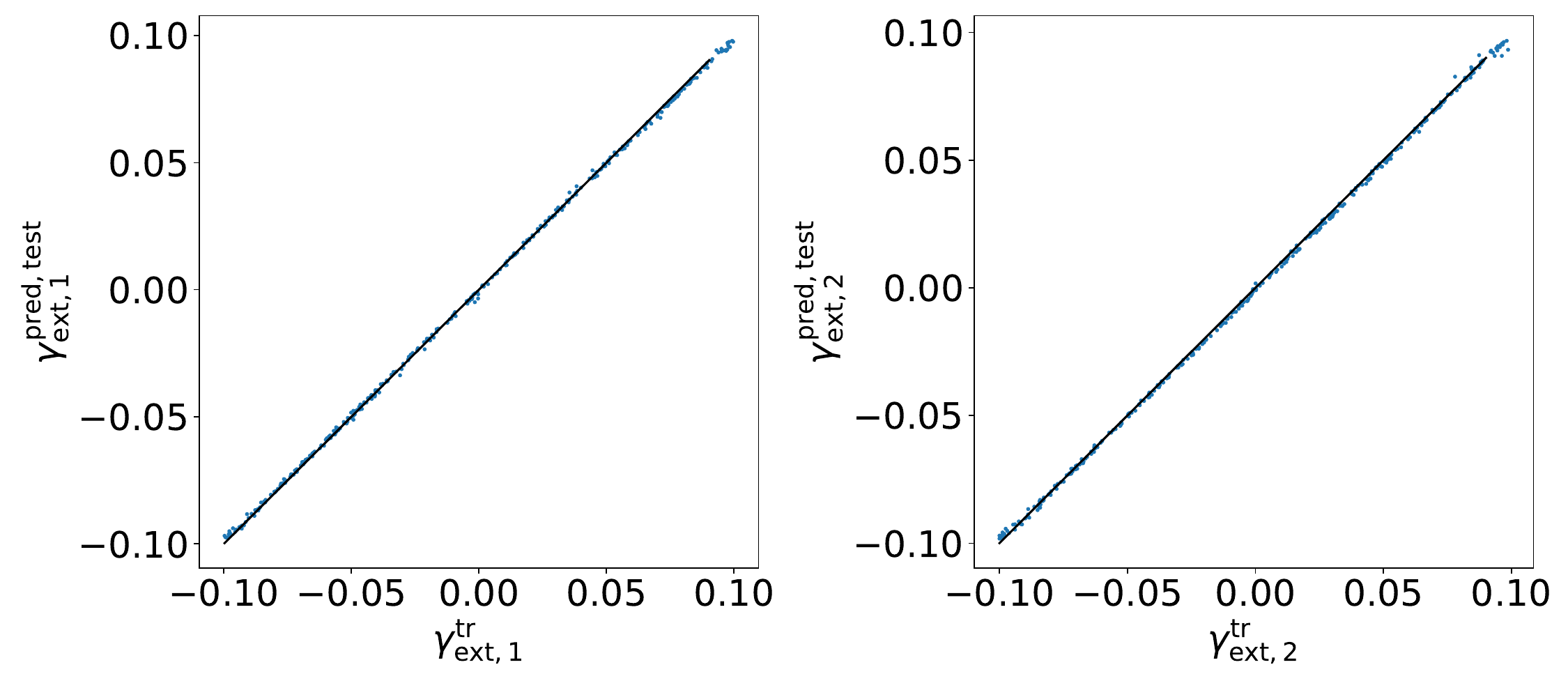}
  \caption{Comparison of ground truth and predictions of the external
shear from the test set by a network trained on 1,000 mocks, each with the same lens and source pair. Under this extreme simplification, the network is able to predict the external shear perfectly even on the new images of the test set.\label{fig:fixpairs} }
\end{centering}
\end{figure}

In the first and second scenarios, the network is able to predict the external shear better but not perfectly. In the third scenario, the network yields a nearly perfect prediction of the external shear on the test data (Fig.~\ref{fig:fixpairs}). This demonstrates the ability of the network to transfer the shear extraction to completely new images. 

As the lens and background source are always the same in the last scenario, we can exclude the possibility that the network connects other features of the image with the external shear parameters. On the other hand, we also exclude possible degeneracies between different parameters, such as the ellipticity and  the external shear, that might explain the difficulties with the external shear. One reason for the good performance in these tests might be the matching PSF for all systems, given that we are using the same lens. Normally, different lenses have slightly different PSFs, meaning that the arcs appear different after the convolution. These variations can introduce difficulties for the network, as it does not know the PSF of a given lens system. Particularly for the external shear, detecting small distortions on the arcs is necessary, which can be highly influenced by different PSF shapes. However, passing the PSFs together with the images to the network did not help (see Sect.~\ref{sec:tests:inputdata}).

\subsection{Variations of the input data}
\label{sec:tests:inputdata}

% here or somewhere else: sqrt stretching
%- test sqrt-squeezing
%  - not really helpful, and since we don't know how it affects the lens image/modeling, we didin't use it further

As   we achieved a better performance in our lens search projects by applying a square-root stretching  \citep{canameras21b,
canameras22a, shu22}, we also tested this for our modeling network. Hence, we no longer passed the images to the network, but rather the square-root of the images after setting all negative background pixels to zero. This helps to enhance the faint arcs compared to the brighter lens in the center, but also changes the flux ratios between pixels on the arcs in any given filter and also the ratio between the different filters where the color information is encoded. Finally, we find no improvement for the modeling network.

% test with subsampling factor, test factor 2: net 101 beter than 104, but no much effect (19031-19060)

Another test of this kind was to subsample the images by a linear interpolation to increase the number of pixels, as our $64\times64$ pixels images are very small compared to images from for example ImageNet. Even though no information was added in this process, the hope was that a higher pixel count might help the network to predict small features and would allow for deeper networks with larger kernel sizes or strides in the convolutional layers. We tested subsampling factors of 2, 3, and 4, but the lowest mean validation loss was obtained without subsampling.

% add FWHM as additional info (20091-20120), not better on ext shear
%    -> maybe network not really able to connect the new input on the FC layers to the images that were already passed through the conv layers

% test with PSF images
% -> does not improve, PSF does not add more info

% train on arcs+PSF images:
% best: 21073**, lens center not as good as before, but can roughly get it
%                ellipticity very hard, not really recovered
%                Einstein radius perfectly recovered
%                external shear similar as before, quite flat distribution, not good recovered.

Given our difficulty in recovering the external shear, we tried to help the network with some additional information. First, we provided the full-width half-maximum (FWHM) values of the point spread function (PSF) frames in addition to the normal set of images. These values were added to the flattened output of the convolutional layers and processed through the FC layers. As this leads to no improvement, we considered networks accepting eight frames instead of four, and added the PSF images directly as input images. To this end, we subsampled the PSF with a linear interpolation to 64 $\times$ 64 pixels, matching the size of the images, and passed them independently of the images through a mirrored branch of convolutional layers, combining the intermediate outputs directly before the FC layers as suggested by \citet{maresca21} and \citet{li22a}. Again, no improvement is seen with this option, possibly because such networks, with their relatively small kernel sizes, perform well in pattern recognition but perhaps not as well in analyzing very similar and completely smooth images like a PSF. Another possible reason for gaining no improvement by adding the PSF images is that these PSFs are likely imperfect, because of the complex stacking procedure of $\sim70$ candidate stars per CCD \citep{bertin11, aihara18a}. These small errors are estimated to reach the $1\%$ level at maximum \citep{aihara18a, aihara19}, but are possibly relevant for the relatively small effects of the external shear on the arcs.

%- test training on arcs alone, net 101-106
%best: 21114**, better than with PSF images, loss -404,31, net 104
%all networks **, so no other one here

To further investigate the external shear, we trained networks on images containing only the arcs. This was proven to be helpful in other modeling networks \citep[e.g.,][]{hezaveh17, levasseur17, morningstar18, morningstar19, pearson21}. By removing the lens light, the remaining information from the arcs becomes more easily accessible. We assumed perfect lens-light subtraction, which is not achievable in reality, but is the best-case scenario when assessing whether or not the network can then better predict the external shear, which is only encoded in the arcs. As expected, the new network predicts the Einstein radius with nearly no scatter over the full range, and also performs well on the lens center. The good performance on the systems with small image separations confirms that the lower performance of our normal network in that regime is due to blending issues with the lens. In this test, we notably last performance on the lens mass ellipticity, which is to some extent expected, as the network retrieves this information partially from the lens \textit{light} which differ slightly from the lens mass distribution. However, also in this case this network cannot accurately predict the external shear, which suggests that the information is very hard to access from ground-based imaging and a generalization over the whole sample is currently impossible.

% test with just one or two filter: G, GR, GI as combinations (20151-20180)
%    -> with one filter slightly worse, but not totally failing => works definetly with Euclid data!
%    -> with two filters very similar -> doesn't make much difference to use 2,3, or 4 filter. Especially Z band is assumed to be not that relevant

To answer the question of the minimum number of filters required, we trained networks on fewer filters. We tried with either only the $g$ band; the $g$ and $r$ bands; or the $g$ and $i$ bands. Although we performed no real optimization of the network architecture and hyper-parameters, we found notably poorer performance with one or two filters than with four. However, even with just one filter, the network is able to predict the SIE parameters reasonably well, and has the greatest difficulty with the ellipticity. This confirms that it is possible to train CNNs or ResNets on single-band images, for example, from Euclid, where the much better resolution will compensate for the missing color information to some extent. As expected, the external shear is not predictable at all with just one filter of HSC image quality within our few tests in this direction.

\FloatBarrier
\section{Image positions and time-delay comparison}
\label{sec:ImPosTimeDelays}

Given the extremely low computational time needed by the trained network to predict the SIE+$\gamma_\text{ext}$ parameters, the network would be ideal for predicting the mass model of a lensing system with a lensed transient. This mass model could then be used to predict, based on the first detection, the next appearing images with corresponding time delays and therefore help to plan follow-up observations. To test the precision and accuracy of the network, we performed a similar test as in \citetalias{schuldt21a} on our test set. To this end, we computed the image positions ($\theta_\text{x}^\text{tr}$, $\theta_\text{y}^\text{tr}$) and time delays $\Delta t ^\text{tr}$ of the background source center given the true mass model from our simulation pipeline. Assuming that the first appearing true image position ($\theta_\text{x,A}^\text{tr}$, $\theta_\text{y,A}^\text{tr}$) is the first observed image and is therefore coincident with ($\theta_\text{x,A}^\text{pred}$, $\theta_\text{y,A}^\text{pred}$), we computed the source position using the mass model predicted by the network. From this predicted source position, we then obtained all other image positions ($\theta_\text{x}^\text{pred}$, $\theta_\text{x}^\text{pred}$) and corresponding time delays $\Delta t ^\text{pred}$. We finally compared the true and predicted values and directly computed the differences for lens systems with a matching number of images.

\begin{figure}[ht!]
\begin{center}
  \includegraphics[trim=0 0 0 0, clip, width=0.5\textwidth]{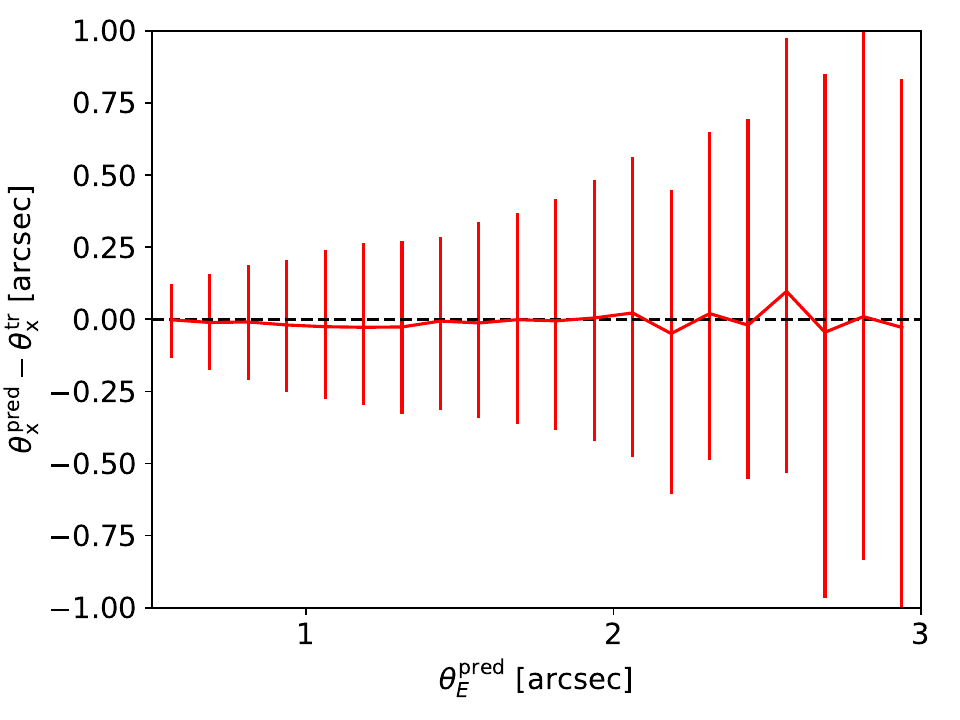}
  \includegraphics[trim=0 0 0 0, clip, width=0.5\textwidth]{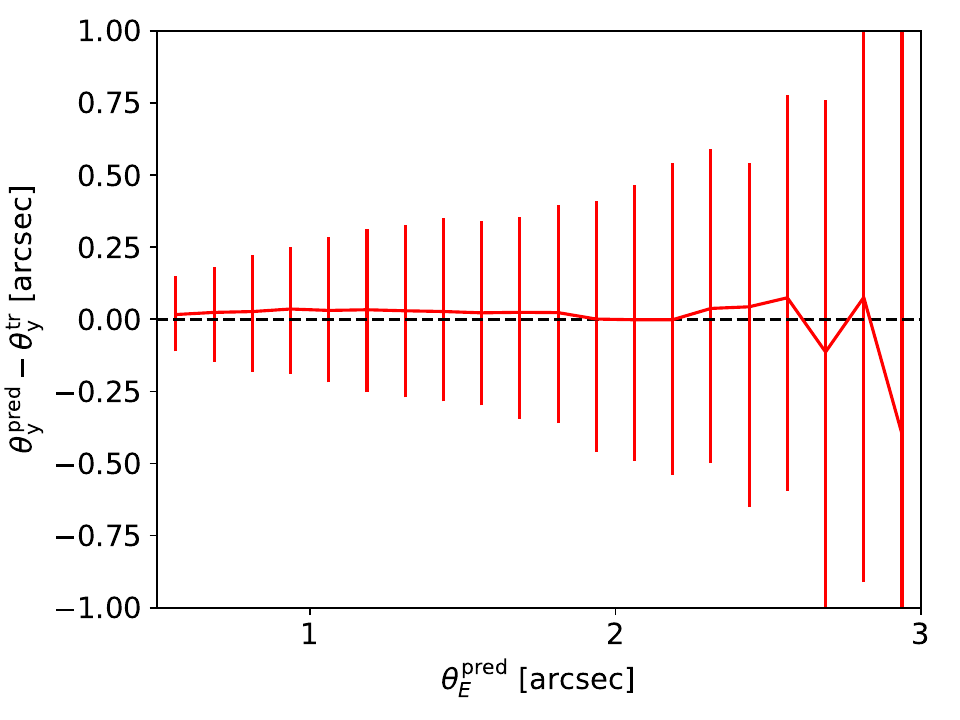}
\caption{Differences between the true and predicted $x$-coordinates (top) and $y$-coordinates (bottom) of the image positions as a function of the predicted Einstein radii.\label{fig:ImPos} }
\end{center}
\end{figure}

To visualize the results, we show the differences of the $x$- and $y$-coordinates of the multiple image positions in Fig.~\ref{fig:ImPos} as a function of the Einstein radius. We find essentially no bias and a similar scatter as in \citetalias{schuldt21a}, which increases with the Einstein radius as expected from Fig.~\ref{fig:comparison}. The small bias and fluctuations in the last bins ($\theta_\text{E} \geq 2\arcsec$) are due to low-number statistics (compare also the histogram of the Einstein radius distribution in Fig.~\ref{fig:comparison}). 

\begin{figure*}[ht!]
\begin{center}
  \includegraphics[trim=0 0 0 0, clip, width=0.5\textwidth]{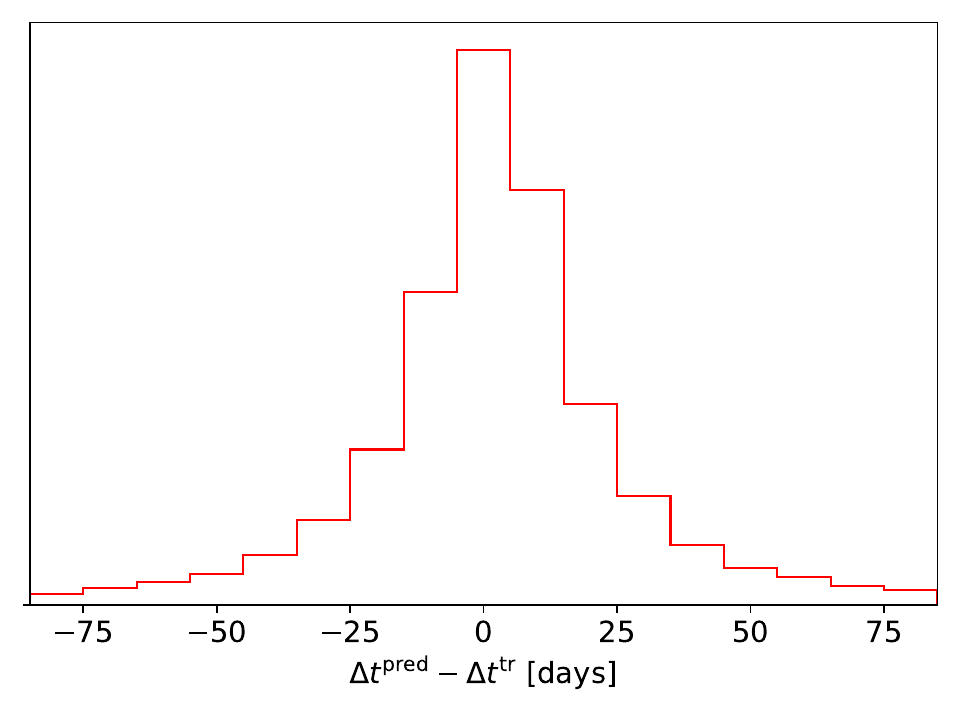}\includegraphics[trim=0 0 0 0, clip, width=0.5\textwidth]{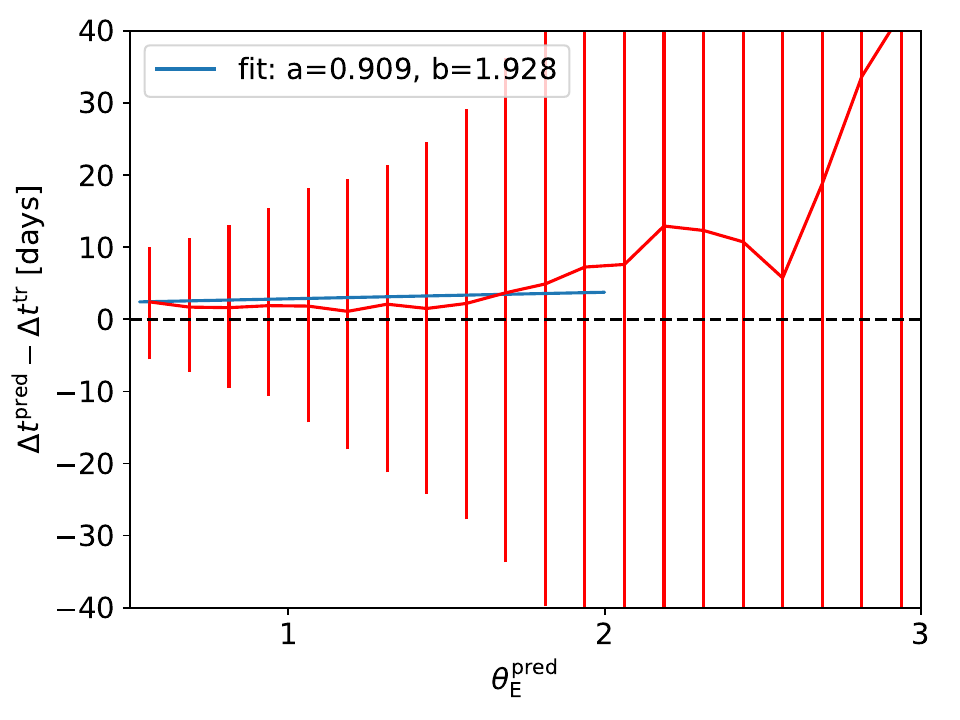}
\caption{Difference between the predicted and true time delays  as a histogram (left) and as a function of the predicted Einstein radii. Given the bias with the Einstein radius, even if relatively small, we fit a linear function $a \times \theta_\text{E}^\text{pred} +b $ in the range $[0.5\arcsec, 2\arcsec]$ as shown in blue.\label{fig:Timedelays} }
\end{center}
\end{figure*}

\begin{figure*}[ht!]
\begin{center}
  \includegraphics[trim=0 0 0 0, clip, width=0.5\textwidth]{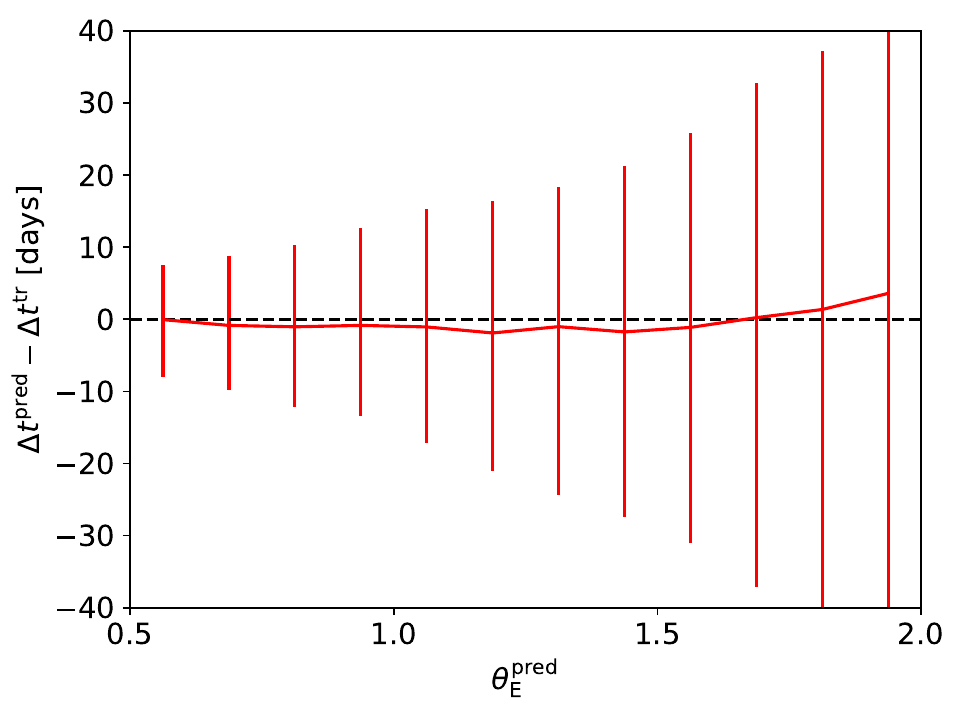}\includegraphics[trim=0 0 0 0, clip, width=0.5\textwidth]{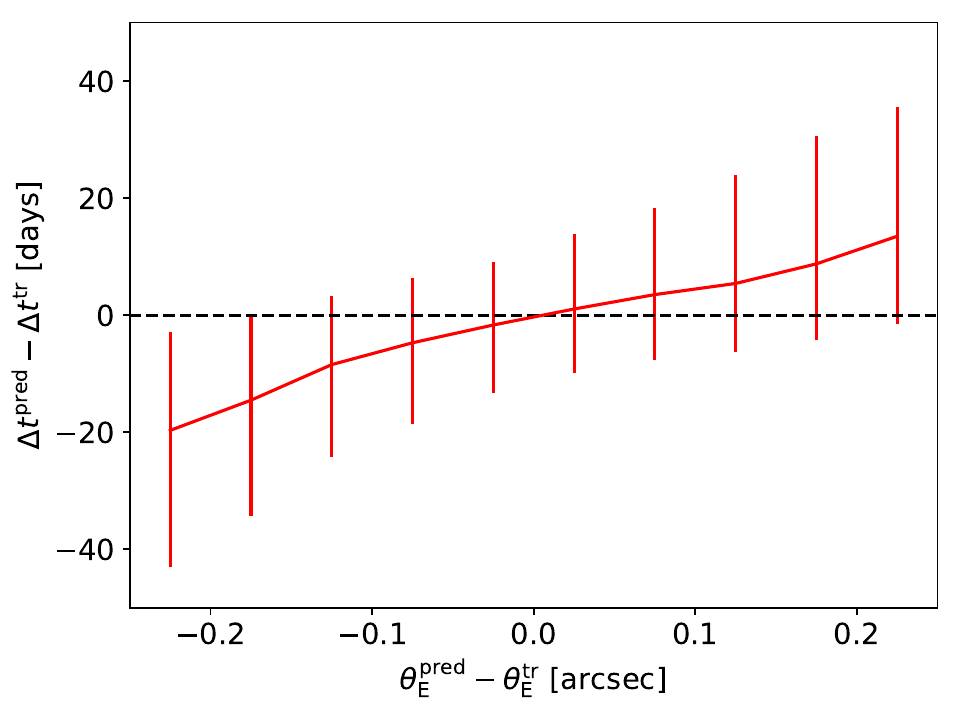}
\caption{Difference between the predicted and true time delays after correction of the linear bias as function of, respectively, the predicted Einstein radius (left panel) and the difference between the predicted and true Einstein radius (right panel).\label{fig:Timedelays_corrected}}
\end{center}
\end{figure*}

Figure~\ref{fig:Timedelays} shows the time-delay differences as a histogram (left panel) and as a function of the predicted Einstein radius (right panel).
%There is a relatively strong bias increasing linearly with the Einstein radius, which is due to $\Delta t$ scaling with $\theta_\text{E}^2$ instead of $\theta_\text{E}$.
We find a small bias in $\Delta t$, while the observed scatter increases as well, which is expected given the performance on the image positions. As we train our ResNet on a sample with equally distributed Einstein radii up to $\sim 1.5\arcsec$, we can compare the bias and scatter to the similarly constructed CNN of \citetalias{schuldt21a} (see Sec.~4.3 in \citetalias{schuldt21a}). These networks show an overall comparable bias towards longer time delays, from which we can infer that the external shear has only a minor effect on the time-delay predictions and that the bias mainly originates from the Einstein radius offsets. While the ResNet  equally under- and overpredicts many time delays with a typical scatter of $\bar{\sigma} \sim 20$ days, the CNN has a slight tendency to make overpredictions. To correct the ResNet predictions for the bias with the Einstein radius, we fit a linear function of the form
\be
a \times \theta_\text{E}^\text{pred} + b
\label{eq:timedelaycorrection}
\ee
to the individual differences, restricting the fitting range to [0.5\arcsec, 2\arcsec]. We obtain a value for $a$ of 0.909 and a value for $b$ of $1.928$. The corrected plot is shown in the left panel of Fig.~\ref{fig:Timedelays_corrected}, along with the corrected time-delay difference $ \Delta t^\text{pred} - \Delta t ^\text{tr}$ as a function of the difference in Einstein radius (right panel). For systems with $\theta_\text{E}^\text{pred} \lesssim 2\arcsec$, we can control the bias within $\sim$ 3 days, although the scatter remains large, limiting the usefulness of the network in predicting precise time delays. This performance is clearly not as good as that of traditional MCMC sampling on ground-based images and a SIE+$\gamma_\text{ext}$ mass distribution, but given the extreme difference in computational time, it may still serve as a reasonable first estimate for systems with $\theta_\text{E} \leq 2\arcsec$.

%\FloatBarrier %HERE
\section{Summary and conclusion}
\label{sec:conclusion}

\citetalias{schuldt21a} demonstrated the possibility of modeling ground-based images of strongly lensed galaxy-scale systems with a relatively simple CNN inspired by the LeNet architecture. Building upon these results, we developed a way of modeling such systems when including the external shear component in addition to the SIE mass distribution of the lens. Moreover, we now predict a 1$\sigma$ uncertainty for each parameter and lensing system. To this end, we make use of a residual neural network, which is a specific type of CNN that includes so-called residual blocks with a skip connection. A diagram of our final network architecture is shown in Fig.~\ref{fig:networksketch}. Because of the included error prediction, we changed the loss function from an MSE loss to a log-probability function with a regression term inspired by \citet{levasseur17}. To ensure that  all parameters contribute equally to the loss, we introduced a scaling of each parameter to the range $[0,1]$ and therefore included the sigmoid function as the last layer.

The network was trained on mock images created from real observed data by only simulating the lensing effect, resulting in realistic mocks, as shown in Fig.~\ref{fig:overview_newmocks}. We used images of LRGs observed with HSC as lenses, together with velocity dispersion and redshift measurements from SDSS. The background sources were images from HUDF with known redshifts.

The training data were created with a flat distribution in the Einstein radius between $0.5\arcsec$ and $\sim 2\arcsec$ and a flat distribution in the external shear strength $\gamma_\text{ext}$ between 0 and 0.1. Moreover, we included Poisson noise on the arcs, which we approximated from the variance map provided by HSC, and improved the lens center and ellipticity estimation using a dedicated mask for each individual lens. To make sure that the network predicts the lens mass center, we applied random shifts to the final mock images of up to three pixels in both $x$ and $y$ directions.

With this procedure, we created $\sim165,000$ mock systems in four filters, which we then used to train our neural network. Through extensive tests, we found a network that is able to accurately predict the SIE parameters of the lens mass profile (Fig.~\ref{fig:comparison} and Fig.~\ref{fig:cornerplot}). However, the external shear  is very hard to predict accurately. We carried out many tests on the external shear, such as up-weighting its loss contribution; only predicting the external shear; adding further information, such as the FWHM values or PSF images; or applying subsampling.
Only in the case of training on a very small sample with always the same lens and source pair, as shown in Fig.~\ref{fig:fixpairs}, is the network  also able to predict the external shear well for new images in the test set. This demonstrates that the network indeed obtained its external shear prediction from the arcs, which are only different because of the variable external shear. In short, the network is able to predict the external shear from effects introduced by the external shear. Therefore, it seems that the network is in general able to extract the information from the external shear, but cannot generalize well to other systems, which is probably a result of the combination of the complexity of the lensing system, image resolution, inaccurate masking of the sources affecting the mocks, the unknown PSF for the network, correlations between the external shear and other parameters, and/or other reasons. This is supported by the fact that the external shear can be relatively well predicted by CNNs trained on more idealized, lens-light subtracted, and high-resolution images \citep{morningstar18}.

In analogy to \citetalias{schuldt21a}, we used the predicted SIE+$\gamma_\text{ext}$ mass parameters to predict the next appearing image(s) and corresponding time delay(s) given the first appearing image of the true, simulated mass model. Although we observed stronger discrepancies on the external shear, the observed scatter on the image positions and time delays is comparable to that obtained with our CNNs of \citetalias{schuldt21a}. We find a bias in the predicted time delays as a function of the predicted Einstein radius, which can be compensated up to $\theta_\text{E}^\text{pred} \sim 2\arcsec$ by applying a linear correction function. Through a comparison of the performance with that of the CNN presented in \citetalias{schuldt21a}, we see that the external shear has only a minor effect on the time-delay prediction.

The greatest strength of our network is certainly its reduced requirement for computational time and its greater degree of automation. Once trained, it predicts these parameters in fractions of a second, while state-of-the-art methods like \GG \, require at least days and a lot of user input for the same task. With this network, we are able to predict the SIE+$\gamma_\text{ext}$ values with uncertainties for all known HSC lenses or lens candidates ---which already number a few thousand--- within minutes. Given the good match of the HSC images to the expected quality of LSST, the performance of our network is expected to hold for LSST as well. Here, we propose to generate dedicated mocks and train a separate network as soon as data from the first LSST data release are available in order to avoid a possible degradation in performance because of slightly different image characteristics.

\FloatBarrier
\begin{acknowledgements}
We thank D.~Sluse and the anonymous referee for helpful comments. SS, SHS, and RC thank the Max Planck Society for support through the
Max Planck Research Group for SHS. This project has received funding
from the European Research Council (ERC) under the European Unions
Horizon 2020 research and innovation programme (LENSNOVA: grant
agreement No 771776).
This research is supported in part by the Excellence Cluster ORIGINS which is funded by the Deutsche Forschungsgemeinschaft (DFG, German Research Foundation) under Germany's Excellence Strategy -- EXC-2094 -- 390783311. YS acknowledges support from the Alexander von Humboldt Foundation in the framework of the Max Planck-Humboldt Research Award endowed by the Federal Ministry of Education and Research.
\\
Based on observations made with the NASA/ESA Hubble Space Telescope, obtained from the data archive at the Space Telescope Science Institute. STScI is operated by the Association of Universities for Research in Astronomy, Inc. under NASA contract NAS 5-26555\\
The Hyper Suprime-Cam (HSC) collaboration includes the astronomical communities of Japan and Taiwan, and Princeton University. The HSC instrumentation and software were developed by the National Astronomical Observatory of Japan (NAOJ), the Kavli Institute for the Physics and Mathematics of the Universe (Kavli IPMU), the University of Tokyo, the High Energy Accelerator Research Organization (KEK), the Academia Sinica Institute for Astronomy and Astrophysics in Taiwan (ASIAA), and Princeton University. Funding was contributed by the FIRST program from Japanese Cabinet Office, the Ministry of Education, Culture, Sports, Science and Technology (MEXT), the Japan Society for the Promotion of Science (JSPS), Japan Science and Technology Agency (JST), the Toray Science Foundation, NAOJ, Kavli IPMU, KEK, ASIAA, and Princeton University. This paper makes use of software developed for the Rubin Observatory Legacy Survey in Space and Time (LSST). We thank the LSST Project for making their code available as free software at http://dm.lsst.org This paper is based in part on data collected at the Subaru Telescope and retrieved from the HSC data archive system, which is operated by Subaru Telescope and Astronomy Data Center (ADC) at National Astronomical Observatory of Japan. Data analysis was in part carried out with the cooperation of Center for Computational Astrophysics (CfCA), National Astronomical Observatory of Japan. We make partly use of the data collected at the Subaru Telescope and retrieved from the HSC data archive system, which is operated by Subaru Telescope and Astronomy Data Center at National Astronomical Observatory of Japan.\\
Funding for the Sloan Digital Sky Survey IV has been provided by the Alfred P. Sloan Foundation, the U.S. Department of Energy Office of Science, and the Participating Institutions. SDSS-IV acknowledges support and resources from the Center for High-Performance Computing at the University of Utah. The SDSS web site is www.sdss.org.\\
SDSS-IV is managed by the Astrophysical Research Consortium for the
Participating Institutions of the SDSS Collaboration including the
Brazilian Participation Group, the Carnegie Institution for Science,
Carnegie Mellon University, the Chilean Participation Group, the
French Participation Group, Harvard-Smithsonian Center for
Astrophysics, Instituto de Astrof\'isica de Canarias, The Johns
Hopkins University, Kavli Institute for the Physics and Mathematics of
the Universe (IPMU) / University of Tokyo, the Korean Participation
Group, Lawrence Berkeley National Laboratory, Leibniz Institut f\"ur
Astrophysik Potsdam (AIP), Max-Planck-Institut f\"ur Astronomie (MPIA
Heidelberg), Max-Planck-Institut f\"ur Astrophysik (MPA Garching),
Max-Planck-Institut f\"ur Extraterrestrische Physik (MPE), National
Astronomical Observatories of China, New Mexico State University, New
York University, University of Notre Dame, Observat\'ario Nacional /
MCTI, The Ohio State University, Pennsylvania State University,
Shanghai Astronomical Observatory, United Kingdom Participation Group,
Universidad Nacional Aut\'onoma de M\'exico, University of Arizona,
University of Colorado Boulder, University of Oxford, University of
Portsmouth, University of Utah, University of Virginia, University of
Washington, University of Wisconsin, Vanderbilt University, and Yale
University. Software Citations: This work uses the following software packages:
\href{https://github.com/astropy/astropy}{\texttt{Astropy}}
\citep{astropy1, astropy2},
\href{https://github.com/matplotlib/matplotlib}{\texttt{matplotlib}}
\citep{matplotlib},
\href{https://github.com/numpy/numpy}{\texttt{NumPy}}
\citep{numpy1, numpy2},
\href{https://www.python.org/}{\texttt{Python}}
\citep{python},
\href{https://github.com/scipy/scipy}{\texttt{Scipy}}
\citep{scipy},
\href{https://pytorch.org}{\texttt{torch}}
\citep{torch}.
\end{acknowledgements}

\bibliographystyle{aa}
\bibliography{NetModel}

\end{document}